\documentclass[conference]{IEEEtran}
\usepackage{array, latexsym, graphicx,times,amsfonts,picture,color}
\usepackage{url}
\usepackage{epsfig}
\usepackage{amsmath}
\usepackage{verbatim}
\usepackage{blkarray}
\usepackage{setspace}
\usepackage{subfigure}
\usepackage{multirow}
\usepackage{breqn}

\def\uzvert{\mbox{$\joinrel{\mid}{\grave{}}\,$}}

\usepackage{txfonts}
\usepackage[T1]{fontenc}
\usepackage{textcomp}

\begin{document}

\title{On the Design of LIL Tests for (Pseudo) Random
Generators and Some Experimental Results}


\makeatletter
\def\ps@headings{%
\def\@oddhead{\mbox{}\scriptsize\rightmark \hfil \thepage}%
\def\@evenhead{\scriptsize\thepage \hfil \leftmark\mbox{}}%
\def\@oddfoot{}%
\def\@evenfoot{}}
\makeatother
\pagestyle{headings}
\hyphenation{op-tical net-works semi-conduc-tor}

\newtheorem{theorem}{Theorem}[section]
\newtheorem{lemma}[theorem]{Lemma}
\newtheorem{corollary}[theorem]{Corollary}
\newtheorem{proposition}[theorem]{Proposition}
\newtheorem{ex}[theorem]{Example}
\newtheorem{definition}[theorem]{Definition}
\newtheorem{fact}[theorem]{Fact}


\author{
\IEEEauthorblockN{Yongge Wang}
\IEEEauthorblockA{Dept. SIS, UNC Charlotte\\
Charlotte, NC 28223, USA\\
Email: yongge.wang@uncc.edu}
}

\maketitle

\newif{\ifbibfile}
\bibfiletrue

\newif{\ifFullversion}

\begin{abstract}
Random numbers have been one of the most useful 
objects in statistics, computer science, cryptography,
modeling, simulation, and other applications
though it is very difficult to construct true randomness.
Many solutions (e.g., cryptographic pseudorandom generators)
have been proposed to harness or simulate randomness
and many statistical testing techniques have been proposed
to determine whether a pseudorandom generator produces
high quality randomness. NIST SP800-22 (2010)  proposes the state 
of art testing suite for (pseudo) random generators
to detect deviations of a binary sequence from randomness.
On the one hand, as a counter example to NIST SP800-22 test suite, 
it is easy to construct functions that are considered
as GOOD pseudorandom generators by  NIST SP800-22
test suite though the output of these functions are easily
distinguishable from the uniform distribution. Thus 
these functions are not  pseudorandom generators by definition.
On the other hand, NIST SP800-22 does not cover some of 
the important laws for randomness. Two fundamental limit theorems
about random binary strings are the central limit 
theorem and the law of the iterated logarithm (LIL). 
Several frequency related tests in NIST SP800-22
cover the central limit theorem while
no NIST SP800-22 test covers LIL.

This paper proposes techniques to address the above challenges that 
NIST SP800-22 testing suite faces.
Firstly, we propose statistical distance 
based testing techniques for (pseudo) random generators
to reduce the above mentioned Type II errors in 
NIST SP800-22 test suite. Secondly, we propose
LIL based statistical testing techniques,
calculate the probabilities, and carry out
experimental tests on widely used pseudorandom
generators by generating around 30TB of pseudorandom sequences. 
The experimental results show that for a sample size of 1000 sequences (2TB), 
the statistical distance between
the generated sequences and the uniform distribution 
is around 0.07 (with $0$ for statistically
indistinguishable and $1$ for completely distinguishable)
and the root-mean-square deviation is around 0.005.
Though the statistical distance 0.07 and RMSD 0.005 are 
acceptable for some applications, for a cryptographic 
``random oracle'', the preferred statistical distance
should be smaller than 0.03 and RMSD be smaller than 0.001 
at the sample size 1000. 
These results justify the importance 
of LIL testing techniques designed in this paper.
The experimental results in this paper are reproducible 
and the raw experimental data are available at author's website.
\end{abstract}

\section{Introduction}
\label{arraybpxorsec}
Secure cryptographic hash functions
such as SHA1, SHA2, and SHA3
and symmetric key block ciphers (e.g., AES and TDES) 
have been commonly used to design pseudorandom generators 
with counter modes (e.g., in Java Crypto Library 
and in NIST SP800-90A standards). 
Though security of hash functions such as SHA1, SHA2, and 
SHA3 has been extensively 
studied from the one-wayness and collision resistant aspects,
there has been limited research on the quality of long pseudorandom 
sequences generated by cryptographic hash functions.
Even if a hash function (e.g., SHA1) performs
like a random function based on existing statistical tests
(e.g., NIST SP800-22 Revision 1A \cite{nist80022}),
when it is called many times for a long sequence generation, 
the resulting long sequence may not satisfy the properties of pseudorandomness
and could be distinguished from a uniformly chosen sequence.
For example, the recent reports from New York Times \cite{nytSnowden}
and The Guardian \cite{guardianSnowden} show that NSA has included 
back doors in NIST SP800-90A pseudorandom bit generators (on which our 
experiments are based on) to get online cryptanalytic capabilities.

Statistical tests are commonly used as a first step in determining 
whether or not a generator produces high quality random bits.
For example, NIST SP800-22 Revision 1A
\cite{nist80022} proposed the state of art statistical
testing techniques for determining whether a random or pseudorandom 
generator is suitable for a particular cryptographic application.
NIST SP800-22  includes 15  tests: frequency (monobit), 
number of 1-runs and 0-runs,  longest-1-runs,
binary matrix rank, discrete Fourier transform, 
template matching, Maurer's ``universal statistical''  test, 
linear complexity,  serial test, 
the approximate entropy, the cumulative sums (cusums),
the random excursions, and the random excursions variants.
In a statistical test of \cite{nist80022},  
a significance level $\alpha\in [0.001, 0.01]$
is chosen for each test.  For each input sequence, a $P$-value 
is calculated and the input string is accepted as pseudorandom 
if $P$-value $\ge\alpha$. A pseudorandom generator is considered
good if, with probability $\alpha$, the sequences produced 
by the generator fail the test. For an in-depth 
analysis, NIST SP800-22 recommends additional 
statistical procedures such as the examination of 
P-value distributions (e.g., using $\chi^2$-test).

NIST SP800-22 test suite has inherent limitations 
with straightforward Type II errors. For example, for a function $F$
that mainly outputs ``random strings'' but, with probability 
$\alpha$, outputs biased strings (e.g., strings consisting mainly of 0's), 
$F$ will be considered as a
``good'' pseudorandom generator by NIST SP800-22 test 
though the output of $F$ could be
distinguished from the uniform distribution (thus, 
$F$ is not a pseudorandom generator by definition). 
In the following, we use two examples to illustrate this kind of Type II errors.
Let $\mbox{RAND}_{c,n}$ be the sets of Kolmogorov $c$-random 
binary strings of length $n$, where $c\geq 1$. 
That is, for a universal Turing machine $M$, let 
\begin{equation}
\label{randcndef}
\mbox{RAND}_{c,n} =
         \left\{x\in\{0,1\}^n: \mbox{ if }M(y)=x\mbox{ then }|y|\geq |x|-c\right\}.
\end{equation}
Let $\alpha$ be a given significance level of NIST SP800-22 test
and ${\cal R}_{2n}={\cal R}_1\cup {\cal R}_2$ where
${\cal R}_1$ is a size $2^n(1-\alpha)$ subset of 
$\mbox{RAND}_{2,2n}$ and ${\cal R}_2$ is a size $2^n\alpha$ subset of 
$\{0^nx: x\in\{0,1\}^{n}\}$. Furthermore, let $f_n:\{0,1\}^n\rightarrow {\cal R}_{2n}$ 
be an ensemble of random functions (not necessarily computable) such 
that $f(x)$ is chosen uniformly at random from ${\cal R}_{2n}$.
Then for each $n$-bit string $x$, with probability $1-\alpha$, 
$f_n(x)$ is Kolmogorov 2-random and with probability $\alpha$, $f_n(x)\in {\cal R}_2$.
Since all Kolmogorov $2$-random strings are guaranteed to pass 
NIST SP800-22 test at significance level $\alpha$ (otherwise, they are 
not Kolmogorov $2$-random by definition) and all strings in 
${\cal R}_2$ fail NIST SP800-22 test at significance level $\alpha$ 
for large enough $n$, the function ensemble  $\{f_n\}_{n\in N}$  is considered 
as a ``good'' pseudorandom generator by NIST SP800-22 test suite.
On the other hand, Theorem 3.2 in  
Wang \cite{wang2002comparison} shows that $\mbox{RAND}_{2,2n} $ (and ${\cal R}_1$)
could be efficiently distinguished from the uniform distribution with a non-negligible 
probability. A similar argument could be used to show that ${\cal R}_{2n}$
could be efficiently distinguished from the uniform distribution with a non-negligible 
probability. In other words,  $\{f_n\}_{n\in N}$ is 
not a cryptographically secure pseudorandom generator. 

As another example, let $\{f'_n\}_{n\in N}$ be a pseudorandom generator
with $f'_n:\{0,1\}^n\rightarrow \{0,1\}^{l(n)}$
where $l(n)>n$. Assume that  $\{f'_n\}_{n\in N}$ is a good
pseudorandom generator by NIST SP800-22 
in-depth statistical analysis of the P-value distributions 
(e.g., using $\chi^2$-test). Define 
a new pseudorandom generators $\{f_n\}_{n\in N}$ as follows:
\begin{equation}
\label{typeIIcounterex}
f_n(x)=\left\{
\begin{array}{ll}
f'_n(x) & \mbox{ if } f'_n(x) \mbox{ contains more 0's than 1's}\\
f'_n(x)\oplus 1^{l(n)} & \mbox{ otherwise }
\end{array}
\right.
\end{equation}
Then it is easy to show  that $\{f_n\}_{n\in N}$ is also a good
pseudorandom generator by NIST SP800-22 
in-depth statistical analysis of the P-value distributions 
(e.g., using $\chi^2$-test). However, the output of  $\{f_n\}_{n\in N}$
is trivially distinguishable from the uniform distribution.

The above two examples show the limitation of
testing approaches specified in NIST SP800-22. The limitation
is mainly due to the fact that NIST SP800-22 does not fully
realize the differences between the two common approaches
to pseudorandomness definitions as observed and analyzed in 
Wang \cite{wang2002comparison}. In other words,
the definition of pseudorandom generators is based on
the indistinguishability concepts though techniques in NIST SP800-22
mainly concentrate on the performance of individual strings.
In this paper, we propose testing techniques that are based on 
statistical distances such as root-mean-square deviation or 
Hellinger distance.  The statistical distance based approach is more accurate 
in deviation detection and avoids above type II errors in 
NIST SP800-22. Our approach is illustrated 
using the LIL test design.

Feller  \cite{feller1945fundamental} mentioned that 
the two fundamental limit theorems of 
random binary strings are the central limit theorem and 
the law of the iterated logarithm. Feller \cite{feller1945fundamental} 
also called attention to the study of the behavior of the maximum of the 
absolute values of the partial sums 
$\bar{S}_n=\frac{\max_{1\le k\le n}|2S(\xi\uzvert k)|-n}{\sqrt{n}}$
and Erdos and Kac \cite{erdos1946certain} obtained the 
limit distribution of $\bar{S}_n$. 
NIST SP800-22 test suite includes several frequency related
tests that cover the first central limit theorem and 
the cusum test, ``the cumulative sums (cusums) test'',
that covers the limit distribution of $\bar{S}_n$. 
However it does not include any test for the important law of the iterated
logarithm. Thus it is important to design LIL based statistical tests.
The law of the iterated logarithm (LIL) says that, 
for a pseudorandom sequence $\xi$, the value $S_{lil}(\xi[0..n-1])$ 
(this value is defined in Theorem \ref{lilili}) 
should stay in $[-1,1]$ and reach both ends infinitely 
often when $n$ increases. It is known  \cite{DBLPWang96,wangThesis,wanginfo5} 
that  polynomial time pseudorandom sequences follow LIL. 
It is also known \cite{feller} that LIL holds for uniform distributions.
Thus LIL should hold for both Kolmogorov complexity based
randomness and for  ``behavioristic'' approach based randomness.

This paper designs LIL based weak, strong, and snapshot 
statistical tests  and obtains formulae for calculating 
the probabilities that 
a random sequence passes the LIL based tests.
We have carried out some experiments to test outcomes 
of several commonly used pseudorandom generators. In particular, we generated 
30TB of sequences using several NIST recommended pseudorandom generators.
Our results show that at the sample size 1000 (or 2TB of data), 
sequences produced by several commonly used
pseudorandom generators have a LIL based statistical distance 0.07
from true random sources. On the other hand, at the sample size 10000 (20TB of data),
sequences produced by NIST-SHA256 based 
pseudorandom generators have a LIL based statistical 
distance 0.02 from true random sources. 
These distances are larger than expected 
for cryptographic applications.

The paper is organized as follows. Section \ref{classicalsec}
introduces notations. Section \ref{lawsforprandom} 
discusses the law of iterated logarithms (LIL).
Section \ref{normalsec} reviews 
the normal approximation to binomial
distributions. Sections \ref{weakdesignsec}, \ref{weakdesignsec2},
and \ref{strongLILdesign} propose weak and strong 
LIL tests. Section \ref{evaRsec} describes the steps
to evaluate a pseudorandom generator. 
Section introduces Snapshot LIL tests.
Section \ref{experisection}
reports experimental results and we conclude 
with Section \ref{conclusionSec}.

\section{Notations and pseudorandom generators}
\label{classicalsec}
In this paper, $N$ and $R^+$ denotes  the set of natural 
numbers (starting from $0$) and the set of non-negative real numbers, respectively.
$\Sigma=\{0,1\}$ is the binary alphabet, 
$\Sigma^*$ is the set of  (finite) binary strings, 
$\Sigma^n$ is the set of
binary strings of length $n$, and $\Sigma^\infty$ is 
the set of infinite 
binary sequences. The length of a string $x$  is 
denoted by $|x|$.
$\lambda$ is the empty string. 
For strings $x,y\in \Sigma^*$, $xy$ is the 
concatenation of $x$ and $y$, 
$x\sqsubseteq y$ denotes that $x$ is an initial segment of $y$.
For a sequence $x\in \Sigma^*\cup\Sigma^\infty$ 
and a natural number $n\ge 0$, 
$x\uzvert n=x[0..n-1]$ denotes
the initial segment of length $n$ of $x$ 
($x\uzvert n=x[0..n-1]=x$ if $|x|\le n$) while
$x[n]$ denotes the $n$th bit of $x$, i.e., 
$x[0..n-1]=x[0]\ldots x[n-1]$. 
For a set {\bf C} of infinite sequences,  
$Prob[{\bf C}]$ denotes the probability
that $\xi\in {\bf C}$ when $\xi$ is chosen by a uniform 
random experiment. 
Martingales are used to describe betting strategies in probability theory.

\begin{definition}
(Ville \cite{ville})
A {\rm martingale} is a function $F: 
\Sigma^*\rightarrow R^+$ such that, for all $x\in \Sigma^*$,
$$F (x)=\frac{F (x1)+F (x0)}{2}.$$
We say that a martingale $F$ {\em succeeds} on a sequence 
$\xi\in\Sigma^\infty$ if  
$\limsup_n F(\xi[0..n-1])=\infty$. 
\end{definition}

The concept of ``effective similarity'' by Goldwasser and 
Micali \cite{goldwasser1984probabilistic} and Yao \cite{yao}
is defined  as follows: 
Let  $X = \{X_n\}_{n\in N}$ and $Y = \{Y_n\}_{n\in N}$
be two probability ensembles  such that each of $X_n$ and
$Y_n$ is a distribution over $\Sigma^n$. 
We say that $X$ and $Y$ are computationally  
(or statistically)  indistinguishable 
if for every feasible algorithm $A$ (or every algorithm $A$),
the total variation difference between $X_n$ and $Y_n$ is 
a negligible function in $n$.

\begin{definition}
\label{defnmartingaleind}
Let $\{X_n\}_{n\in N}$ and $\{Y_n\}_{n\in N}$
be two probability ensembles.
$\{X_n\}_{n\in N}$ and $\{Y_n\}_{n\in N}$
are computationally  (respectively, statistically)  indistinguishable if for 
any polynomial time computable set $D\in \Sigma^*$ 
(respectively, any set $D\in \Sigma^*$)
and any polynomial $p$,
the  inequality (\ref{ind1})  holds for almost all $n$.
\begin{equation}
\label{ind1}
|Prob[A(X_n)= 1]-Prob[A(Y_n)=1]|\le \frac{1}{p(n)}
\end{equation}
\end{definition}

Let $l : N\rightarrow N$ with $l(n) \ge n$ for all $n\in N$ and 
$G$ be a polynomial-time computable algorithm such that 
$|G(x)|=l(|x|)$ for all $x \in \Sigma^*$.

Then the pseudorandom generator concept \cite{blum,yao} is defined
as follows.

\begin{definition}
\label{defnpseudgene}
Let $l : N\rightarrow N$ with $l(n) > n$ for all $n\in N$,
and $\{U_n\}_{n\in N}$ be the uniform distribution.
A pseudorandom generator is a polynomial-time algorithm $G$ with
the following properties:
\begin{enumerate}
\item $|G(x)|=l(|x|)$ for all $x \in \Sigma^*$.
\item The ensembles $\left\{G(U_n)\right\}_{n\in N}$ and
$\{U_n\}_{n\in N}$ are computationally  indistinguishable.
\end{enumerate}
\end{definition}

Let $\mbox{RAND}_{c}=\cup _{n\in \mathcal{N}}\mbox{RAND}_{c,n}$ 
where  $\mbox{RAND}_{c,n}$  is the set of 
Kolmogorov $c$-random sequences that is defined in 
equation (\ref{randcndef}). 
Then we have
\begin{theorem}
\label{wangcomparithm}
(\cite[Theorem 3.2]{wang2002comparison})
 The ensemble $R_{c}=\{R_{c,n}\}_{n\in \mathcal{N}}$ is not
pseudorandom.
\end{theorem}

Theorem \ref{wangcomparithm} shows the importance for 
a good pseudorandom generator to fail each statistical test
with certain given probability.  

\section{Stochastic Properties of Long Pseudorandom Sequences}
\label{lawsforprandom}
Classical infinite random sequences were first introduced 
as a type of disordered sequences, called ``Kollektivs", 
by von Mises \cite{mises} as a
foundation for probability theory. 
The two  features
characterizing a Kollektiv are: the existence of 
limiting relative
frequencies within the sequence and the invariance of 
these  limits under the operation of an ``{a}dmissible place
selection". Here an admissible place selection is 
a procedure for selecting a
subsequence of a  given sequence $\xi$ in such a 
way that the decision to select
a term $\xi[n]$  does not depend on the value of 
$\xi[n]$. 
Ville \cite{ville} showed that von Mises' approach is not satisfactory 
by proving that: for each countable
set of ``{a}dmissible place selection" rules, there 
exists a ``Kollektiv"  which does not satisfy the 
law of the iterated logarithm (LIL).
Later, Martin-L\"{o}f  \cite{martin} developed the notion of 
random sequences based on the notion of 
typicalness. A sequence is typical if 
it is not in any {\it constructive} null sets. 
Schnorr \cite{schnorr1}
introduced $p$-randomness concepts by defining the 
{\it constructive} null sets as polynomial time 
computable measure $0$ sets.
The law of the iterated logarithm (LIL)
plays a central role in the study of the Wiener process 
and Wang \cite{wanginfo5} showed that LIL holds
for $p$-random sequences.

Computational complexity based pseudorandom sequences have
been studied extensively in the literature. For example, $p$-random
sequences are defined by taking 
each polynomial time computable martingale 
as a statistical test.
\begin{definition}
(Schnorr \cite{schnorr1})
An infinite sequence $\xi\in\Sigma^\infty$ is p-random (polynomial time random)
if for any polynomial time computable martingale $F$, $F$ does not succeed on 
$\xi$.
\end{definition}

A sequence $\xi\in\Sigma^\infty$ is Turing machine computable if there exists 
a Turing machine $M$ to calculate the bits $\xi[0]$, $\xi[1]$, $\cdots$. In the following,
we prove a theorem which says that, for each Turing 
machine computable non $p$-random sequence $\xi$,
there exists a martingale $F$ such that 
the process of $F$ succeeding on $\xi$ can be efficiently observed in  time $O(n^2)$.
The theorem is useful in the characterizations of $p$-random sequences
and in the characterization of LIL-test waiting period. 

\begin{theorem}
\label{wangprandom}
(\cite{wanginfo5})
For a sequence $\xi\in\Sigma^\infty$
and a polynomial time computable martingale $F$, $F$ succeeds on $\xi$ 
if and only if there exists a martingale $F'$ and a non-decreasing 
$O(n^2)$-time computable (with respect to the unary representation of numbers)
function from $h:N\rightarrow N$ such that 
$F'(\xi[0..n-1])\ge h(n)$ for all $n$.
\end{theorem}

 It is shown in \cite{wanginfo5} that $p$-random sequences 
are stochastic in the sense of von Mises and satisfy common
statistical laws such as the law of the iterated logarithm. 
It is not difficult to show 
that all $p$-random sequences pass the NIST SP800-22 \cite{nist80022} 
tests for significance level $0.01$ since each test in \cite{nist80022} 
could be converted to a polynomial time computable martingale
which succeeds on all sequences that do not pass this test. 
However,  none of the sequences generated by
pseudorandom generators are $p$-random since from the generator
algorithm itself, a martingale can 
be constructed to succeed on sequences that it generates. 

Since there is no efficient mechanism to generate $p$-random sequences,
pseudorandom generators are commonly used to produce long
sequences for cryptographic applications.
While the required uniformity property (see NIST SP800-22 \cite{nist80022})
 for pseudorandom sequences
is equivalent to the law of large numbers, 
the scalability property (see \cite{nist80022}) is equivalent to 
the invariance property under the operation of ``admissible 
place selection'' rules. 
Since $p$-random sequences satisfy common statistical laws,
it is reasonable to expect that pseudorandom sequences 
produced by pseudorandom generators  satisfy these laws also (see, e.g.,
\cite{nist80022}). 

The law of the iterated logarithm (LIL) 
describes the fluctuation scales of a random walk.
For a nonempty string $x\in \Sigma^*$, let 
$$S (x)=\sum_{i=0}^{|x|-1}x[i]
\quad\mbox{ and }\quad
S^* (x)=\frac{2\cdot S (x)-|x|}{\sqrt{|x|}}$$
where $S(x)$ denotes the {\it number} of 1s in $x$ and 
$S^*(x)$ denotes the {\it reduced number} of 1s in $x$.
$S^*(x)$ amounts to measuring the deviations of $S(x)$ from
$\frac{|x|}{2}$  in units of $\frac{1}{2} \sqrt{|x|}$.

The law of large numbers says that, for a pseudo random
sequence $\xi$, the limit of $\frac{S(\xi[0..n-1])}{n}$ is $\frac{1}{2}$,
which corresponds to the frequency (Monobit) test in 
NIST SP800-22 \cite{nist80022}.
But it says nothing about the reduced deviation 
$S^*(\xi[0..n-1])$. It is intuitively clear that, for a pseudorandom sequence
$\xi$, $S^*(\xi[0..n-1])$ will sooner or later take on arbitrary large 
values (though slowly). The law of the iterated logarithm (LIL), which 
was first discovered by Khintchine \cite{khintchine1924satz}, 
gives an optimal upper bound $\sqrt{2\ln\ln n}$
for the fluctuations of $S^*(\xi[0..n-1])$. It was showed in Wang \cite{wanginfo5}
that this law holds for $p$-random sequences also.

\begin{theorem}
\label{lilili}
(LIL for $p$-random sequences \cite{wanginfo5})
For a sequence $\xi\in\Sigma^\infty$, let 
\begin{equation}
\label{lildefn}
S_{lil}(\xi\uzvert n)= \frac{2\sum_{i=0}^{n-1}\xi [i] -n}{\sqrt{2n\ln\ln n}}
\end{equation}
Then for each  $p$-random sequence $\xi\in\Sigma^\infty$
we have  both
$$
\limsup_{n\rightarrow\infty}S_{lil}(\xi\uzvert n)=1
\mbox{ and }
\liminf_{n\rightarrow\infty}S_{lil}(\xi\uzvert n)=-1.$$
\end{theorem}

\section{Normal Approximations to $S_{lil}$}
\label{normalsec}
In this section, we provide several results on normal approximations to
the function $S_{lil}(\cdot)$ that will be used in next sections. 
The DeMoivre-Laplace theorem is a normal approximation to 
the binomial distribution, which says that 
the number of ``successes'' in $n$ independent coin flips 
with head probability $1/2$ is approximately a 
normal distribution with mean $n/2$ and standard deviation $\sqrt{n}/2$.
We first review a few classical results on the normal 
approximation to the binomial distribution.

\begin{definition}
The normal density function with mean $\mu$ and variance $\sigma$ 
is defined as
\begin{equation}
\label{normaldensityfungen}
f(x)=\frac{1}{\sigma\sqrt{2\pi}}e^{-\frac{(x-\mu)^2}{2\sigma^2}};
\end{equation} 
For $\mu=0$ and $\sigma=1$, we have the standard 
normal density function
\begin{equation}
\label{normaldensityfun}
\varphi(x)=\frac{1}{\sqrt{2\pi}}e^{-\frac{x^2}{2}},
\end{equation}
its integral 
\begin{equation}
\label{normaldistributionfunc}
\Phi(x)=\int_{-\infty}^x\varphi(y)dy
\end{equation}
is the standard normal distribution function.
\end{definition}

\begin{lemma}
(\cite[Chapter VII.1, p175]{feller}) 
For every $x>0$, we have 
\begin{equation}
\label{normalappro}
(x^{-1}-x^{-3})\varphi(x)<1-\Phi(x)<x^{-1}\varphi(x)
\end{equation}
\end{lemma}

The following DeMoivre-Laplace limit theorem is derived  
from the approximation Theorem on page 181 of \cite{feller}.

\begin{theorem}
\label{approximation}
For fixed $x_1, x_2$, we have 
\begin{equation}
\label{normalapprothm}
\lim_{n\rightarrow \infty }
Prob\left[x_1\le S^*(\xi\uzvert n)\le x_2\right] =  \Phi(x_2)-\Phi(x_1).
\end{equation}
The growth speed for the above approximation is bounded by $\max\{k^2/n^2, k^4/n^3\}$
where $k=S(\xi\uzvert n)-\frac{n}{2}$.
\end{theorem}

The following lemma is useful for interpreting $S^*$ 
based approximation results 
into $S_{lil}$ based approximation. It is obtained by noting the fact that
$\sqrt{2\ln\ln n}\cdot S_{lil}(\xi\uzvert n)=S^*(\xi\uzvert n)$.
\begin{lemma}
\label{dlaplaceslil} 
For any $x_1,x_2$, we have 
\begingroup
\def\arraystretch{1.4}
$$\begin{array}{l}
Prob\left[x_1<S_{lil}(\xi\uzvert n)<x_2\right]\\
\quad = Prob\left[x_1\sqrt{2\ln\ln n}<S^*(\xi\uzvert n)<x_2\sqrt{2\ln\ln n}\right]
\end{array}$$
\endgroup
\end{lemma}

In this paper, we only consider tests for $n\ge 2^{26}$ and $x_2\le 1$. 
That is, $S^*(\xi\uzvert n)\le \sqrt{2\ln\ln n}$.
Thus 
$$k=S(\xi\uzvert n)-\frac{n}{2}\simeq \frac{\sqrt{n}}{2}S^*(\xi\uzvert n)\le
\sqrt{2n\ln\ln n}/2.
$$ 
Hence, we have 
$$\max\left\{\frac{k^2}{n^2}, \frac{k^4}{n^3}\right\}=\frac{k^2}{n^2}
=\frac{(1-\alpha)^2\ln\ln n}{2n}< 2^{-22}
$$
By Theorem  \ref{approximation}, the approximation probability 
calculation errors in this paper will be less than $0.0000002<2^{22}$
which is negligible. Unless stated otherwise, we will not mention 
the approximation errors in the remainder of this paper.

\section{Weak-LIL test and design}
\label{weakdesignsec}
Theorem \ref{lilili} shows that pseudorandom
sequences should satisfy the law of the iterated logarithm (LIL).
Thus we propose the following weak LIL test for random sequences.

\noindent
{\bf Weak LIL Test}:
Let $\alpha\in (0, 0.25]$ and $\aleph\subset N$
be a subset of natural numbers, 
we say that a sequence $\xi$ does not pass the weak
$(\alpha,\aleph$)-LIL test if $-1+\alpha<S_{lil}(\xi\uzvert n)<1-\alpha$ 
for all $n\in \aleph$. Furthermore,  ${\bf P}_{(\alpha, \aleph)}$ 
denotes the probability that a random sequence passes the weak
$(\alpha,\aleph)$-LIL test, and  ${\bf E}_{(\alpha, \aleph)}$ is the set
of sequences that pass the weak $(\alpha,\aleph)$-LIL test.

By the definition, 
a sequence $\xi$ passes the weak $(\alpha,\aleph$)-LIL test
if $S_{lil}$ reaches either $1-\alpha$ or $-1+\alpha$
at some points in  $\aleph$.
In practice, it is important to 
choose appropriate test point set $\aleph$ and 
calculate the probability for a random
sequence $\xi$ to pass the weak $(\alpha, \aleph)$-LIL test. 
In this section we calculate the probability for a sequence
to pass the weak $(\alpha, \aleph)$-LIL 
test with the following choices of $\aleph$:
$$\aleph_0=\{2^{0}n_1\}, \cdots, \aleph_t=\{2^{t}n_1\}, 
\mbox{ and } \bigcup\aleph_i$$
for  given $n_1$ and $t$. Specifically, we will consider the 
cases for $t=8$ and $n_1=2^{26}$.

\begin{theorem}
\label{sliladd}
Let $x_1,\cdots, x_t\in \{0,1\}^n$. Then we have 
\begin{equation}
\label{liladd}
S_{lil}(x_1)+\cdots +S_{lil}(x_t)=
S_{lil}(x_1\cdots x_t)\cdot \sqrt{\frac{t\ln\ln (tn)}{\ln\ln n}}
\end{equation}
\end{theorem}

\noindent
{\em Proof.} By (\ref{lildefn}), we have
\begin{dmath}
S_{lil}(x_1)+\cdots +S_{lil}(x_t)
=\displaystyle\frac{2\sum_{i=1}^tS(x_i)-tn}{\sqrt{2n\ln\ln n}}
=\displaystyle\frac{2\cdot S(x_1\cdots x_t)-tn}{\sqrt{2n\ln\ln n}}
= \displaystyle\frac{2\cdot S(x_1\cdots x_t)-tn}
  {\sqrt{2\cdot tn\ln\ln tn}}
 \cdot  \sqrt{\displaystyle\frac{t\ln\ln tn}{\ln\ln n}}
=S_{lil}(x_1\cdots x_t)\cdot \sqrt{\displaystyle\frac{t\ln\ln (tn)}{\ln\ln n}}
\end{dmath}
\hfill{$\Box$}

Theorem \ref{sliladd} can be generalized as follows.
\begin{theorem}
\label{sliladdVar}
Let $x_1\in  \{0,1\}^{sn}$ and $x_2\in \{0,1\}^{tn}$. Then we have 
\begin{dmath}
\label{liladdvar}
S_{lil}(x_1)\sqrt{s\ln\ln (sn)} +S_{lil}(x_2)\sqrt{t\ln\ln (tn)}
=S_{lil}(x_1x_2)\sqrt{(s+t)\ln\ln ((s+t)n)}
\end{dmath}
\end{theorem}

\noindent
{\em Proof.} We first note that 
\begin{equation}
\label{rr4}
S_{lil}(x_1)\sqrt{s\ln\ln (sn)}=(2\cdot S(x_1)-sn)/\sqrt{2n}
\end{equation}
\begin{equation}
\label{rr5}
S_{lil}(x_2)\sqrt{t\ln\ln (tn)}=(2\cdot S(x_2)-tn)/\sqrt{2n}
\end{equation}
By adding equations (\ref{rr4}) and (\ref{rr5}) together,
we get (\ref{liladdvar}). The theorem is proved.
\hfill{$\Box$}

\begin{corollary}
\label{corn2nprob}
Let $0<\theta<1$ and $1\le s<t$. For given $\xi\uzvert sn$ with 
$S_{lil}(\xi\uzvert sn)=\varepsilon$ and randomly chosen 
$\xi[sn..tn-1]$, 
\begin{equation}
\label{coroeee2}
\begin{array}{l}
Prob\left[S_{lil}(\xi\uzvert tn)\ge \theta\right]=\\
\quad Prob\left[
S^*(\xi[sn..tn-1])\ge \sqrt{\frac{2}{t-s}}\left(
\theta\sqrt{t\ln\ln tn} -\right.\right.\\
\quad\quad\quad\quad\quad\quad\quad\quad\quad\quad\quad\quad \quad\quad\quad 
\left.\left. 
\varepsilon\sqrt{s\ln\ln sn}\right.\right]
\end{array}
\end{equation}
\end{corollary}

\noindent
{\em Proof.} By Theorem \ref{sliladdVar}, we have  
\begin{equation}
\begingroup
\def\arraystretch{1.4}
\begin{array}{l}
S_{lil}(\xi[0..tn-1])\sqrt{t\ln\ln tn} =\\
\quad S_{lil}(\xi[sn..tn-1])\sqrt{(t-s)\ln\ln (t-s)n} +\varepsilon 
\sqrt{s \ln\ln sn}.
\end{array}
\endgroup
\end{equation}

Thus $S_{lil}(\xi[0..tn-1])\ge \theta$ if, and only if, 
\begin{equation}
\label{equntn}
S_{lil}(\xi[sn..tn-1])\ge
\frac{\theta\sqrt{t\ln\ln tn} -\varepsilon\sqrt{s\ln\ln sn}}
{\sqrt{(t-s)\ln\ln (t-s)n}}
\end{equation}
By Lemma \ref{dlaplaceslil}, (\ref{equntn}) is equivalent to
(\ref{equn2n1t}).
\begin{equation}
\label{equn2n1t}
S^*(\xi[sn..tn-1])\ge \sqrt{\frac{2}{t-s}}
\left(\theta\sqrt{t\ln\ln tn} -\varepsilon\sqrt{s\ln\ln sn}\right)
\end{equation}
In other words, (\ref{coroeee2}) holds. 
\hfill{$\Box$}

After these preliminary results, we will begin 
to calculate the probability for a random sequence 
to pass the weak $(\alpha, \aleph)$-LIL test.

\begin{table*}[ht]
\caption{Weak $(0.1,\aleph)$-LIL and  $(0.05,\aleph)$-LIL test probabilities}
\label{weakLIL01}
\begin{center}
{
\scriptsize
\begin{tabular}{|c|c|c|c|c|c|c|c|c|c|} \hline
$\alpha$ &$\aleph_0$ & $\aleph_1$ & $\aleph_2$ & $\aleph_3$ & $\aleph_4$ & 
    $\aleph_5$& $\aleph_6$ & $\aleph_7$ & $\aleph_8$ \\ \hline
\multicolumn{10}{|c|}{$\alpha=0.1$}\\ \hline
$\aleph_0$&0.03044&0.05085&0.05441&0.05540&0.05544&0.05507&0.05453&0.05394&0.05334\\ \hline
$\aleph_1$&&0.02938&0.04918&0.05263&0.05361&0.05365&0.05331&0.05281&0.05226\\ \hline
$\aleph_2$&&&0.02838&0.04762&0.05097&0.05193&0.05199&0.05168&0.05121\\ \hline
$\aleph_3$&&&&0.02746&0.04616&0.04942&0.05036&0.05043&0.05014\\ \hline
$\aleph_4$&&&&&0.02661&0.04479&0.04797&0.04888&0.04897\\ \hline
$\aleph_5$&&&&&&0.02580&0.04351&0.04660&0.04750\\ \hline
$\aleph_6$&&&&&&&0.02505&0.04230&0.04531\\ \hline
$\aleph_7$&&&&&&&&0.02434&0.04116\\ \hline
$\aleph_8$&&&&&&&&&0.02367\\ \hline
\multicolumn{10}{|c|}{$\alpha=0.05$}\\ \hline
$\aleph_0$&0.02234&0.03770&0.04016&0.04074&0.04065&0.04029&0.03983&0.03935&0.03886\\ \hline
$\aleph_1$&&0.02148&0.03633&0.03871&0.03928&0.03921&0.03888&0.03845&0.03799\\ \hline
$\aleph_2$&&&0.02068&0.03506&0.03737&0.03792&0.03786&0.03756&0.03716\\ \hline
$\aleph_3$&&&&0.01995&0.03387&0.03611&0.03666&0.03661&0.03632\\ \hline
$\aleph_4$&&&&&0.01926&0.03277&0.03494&0.03547&0.03544\\ \hline
$\aleph_5$&&&&&&0.01862&0.03173&0.03384&0.03437\\ \hline
$\aleph_6$&&&&&&&0.01802&0.03076&0.03281\\ \hline
$\aleph_7$&&&&&&&&0.01746&0.02985\\ \hline
$\aleph_8$&&&&&&&&&0.01693\\ \hline
\end{tabular}
}
\end{center}
\end{table*}

\begin{ex}
For $\alpha=0.1$,  $\alpha=0.05$,  and $\aleph_i=\{2^{i+26}\}$ with 
$0\le i\le 8$, the entry at $(\aleph_i, \aleph_i)$ in
Table \ref{weakLIL01} list the probability
${\bf P}_{(\alpha, \aleph)}$ 
that a random sequence passes the weak $(\alpha,\aleph_i)$-LIL test.
\end{ex}

\noindent
{\em Proof.} Let $\theta=1-\alpha$. By Theorem \ref{approximation} and 
Lemma \ref{dlaplaceslil}, 
\begin{dmath}
\label{apppp}
Prob\left[|S_{lil}(\xi\uzvert n)|\ge \theta \right]\simeq
2(1-\Phi(\theta\sqrt{2\ln\ln n})).
\end{dmath}
By substituting $\theta=0.95$ (respectively $0.9$), 
and $n=2^{26}, \cdots, n=2^{34}$
into (\ref{apppp}), we obtain the value ${\bf P}_{(0.1, \aleph_i)}$ 
(respectively  ${\bf P}_{(0.05, \aleph_i)}$) at 
the entry  $(\aleph_i, \aleph_i)$  
in Table \ref{weakLIL01}. This completes 
the proof of the Theorem.
\hfill$\Box$

Now we consider the probability for a random sequence
to pass the weak $(\alpha,\aleph)$-LIL test with 
$\aleph$ as the union of two $\aleph_i$. First we present the following
union theorem.

\begin{theorem}
\label{thmtwoNvee}
For fixed $0<\alpha<1$ and $t\ge 2$, let
$\theta=1-\alpha$, $\aleph=\{n, tn\}$, $\aleph_a=\{n\}$,
 $\aleph_b=\{tn\}$. We have
\begin{equation}
\label{twoNvee}
\begingroup
\def\arraystretch{1.4} 
\begin{array}{l}
{\bf P}_{(\alpha, \aleph)}\simeq
{\bf P}_{(\alpha,\aleph_a)}+\\ \quad
\displaystyle\frac{1}{\pi}\displaystyle\int_{-\theta\sqrt{2\ln\ln n}}^{\theta\sqrt{2\ln\ln n}} 
\displaystyle\int_{\sqrt{\frac{1}{t-1}}(\theta\sqrt{2t\ln\ln tn}-y)}^\infty e^{-\frac{x^2+y^2}{2}} dxdy
\end{array}
\endgroup
\end{equation}
Alternatively, we have 
\begin{equation}
\label{twoNvee2} 
\begingroup
\def\arraystretch{1.4}
\begin{array}{l}
{\bf P}_{(\alpha, \aleph)}
\simeq {\bf P}_{(\alpha,\aleph_a)}+{\bf P}_{(\alpha,\aleph_b)} -\\ \quad
\displaystyle\frac{1}{\pi}\displaystyle\int_{\theta\sqrt{2\ln\ln n}}^{\infty}
\displaystyle\int_{\sqrt{\frac{1}{t-1}}(\theta\sqrt{2t\ln\ln tn}-y)}^\infty e^{-\frac{x^2+y^2}{2}} dxdy
\end{array}
\endgroup
\end{equation}
\end{theorem}

\noindent
{\em Proof.} Since ${\bf E}_{(\alpha,\aleph)}={\bf E}_{(\alpha,\aleph_a)}\cup 
{\bf E}_{(\alpha,\aleph_b)},$ we have 
$${\bf P}_{(\alpha,\aleph)}=\left({\bf P}_{(\alpha,\aleph_a)}+ 
{\bf P}_{(\alpha,\aleph_b)}\right)-{\bf P}_{(\alpha,\aleph_a\cap \aleph_b)}
$$
where 
$${\bf E}_{(\alpha, \aleph_a\cap \aleph_b)}=
\left\{\xi: \left|S_{lil}(\xi\uzvert n)\right|>\theta \bigvee \left|S_{lil}(\xi\uzvert tn)\right|>\theta\right\}.
$$
By symmetry, it suffices to show that 
\begin{equation}
\label{probOF2n}
\begingroup
\def\arraystretch{2}
\begin{array}{l}
Prob\left[S_{lil}(\xi\uzvert tn)\ge \theta 
 | \overline{{\bf E}_{(\alpha,\aleph_a)}}\right]\\\quad
\simeq\frac{1}{2\pi}\int_{-\theta\sqrt{2\ln\ln n}}^{\theta\sqrt{2\ln\ln n}}
\int_{\sqrt{\frac{1}{t-1}}(\theta\sqrt{2t\ln\ln tn}-y)}^\infty e^{-\frac{x^2+y^2}{2}} dxdy
\end{array}
\endgroup
\end{equation}
Let $\Delta_1 =\sqrt{2\ln\ln n}\cdot \Delta z$.
By Corollary \ref{corn2nprob}, the probability that 
$S_{lil}(\xi\uzvert n)\in [z,z+\Delta z]$ 
and $S_{lil}(\xi\uzvert tn)>\theta$
is approximately
\begin{equation}
\label{initiiiii}
\begingroup
\def\arraystretch{2}
\begin{array}{l}
\displaystyle\int_{z\sqrt{2\ln\ln n}}^{z\sqrt{2\ln\ln n}+\Delta_1}\varphi(x)dx
\displaystyle\int^{\infty}_{\sqrt{\frac{2}{t-1}}(\theta\sqrt{t\ln\ln tn}-z\sqrt{\ln\ln n})}\varphi(x)dx  \\
\simeq \Delta_1 \cdot \varphi(z\sqrt{2\ln\ln n}) \cdot
\displaystyle\int^{\infty}_{\sqrt{\frac{2}{t-1}}(\theta\sqrt{t\ln\ln tn}-z\sqrt{\ln\ln n})}\varphi(x)dx
\end{array}
\endgroup
\end{equation}
By substituting $y=z\sqrt{2\ln\ln n}$ and 
integrating the equation (\ref{initiiiii}) over the interval
$y\in [-\theta\sqrt{2\ln\ln n}, \theta\sqrt{2\ln\ln n}]$, we get the equation
(\ref{probOF2n}). 

The equation (\ref{twoNvee2}) could be proved similarly by the following
observation: a sequence passes the weak $(\alpha, \aleph)$-LIL test
if it passes the weak LIL test at point $n$ or at point $2n$. Thus the total
probability is the sum of these two probabilities minus
the probability that the sequence passes the weak LIL test 
at both points at the same time. The theorem is then proved.
\hfill{$\Box$}

\begin{ex}
For $\alpha=0.1$ (respectively $\alpha=0.05$) and 
$\aleph_i=\{2^{i+26}\}$ with 
$0\le i<j\le 8$,  the entry at $(\aleph_i, \aleph_j)$ in 
Table \ref{weakLIL01}
is the probability that a random sequence passes the weak 
$(0.1,\aleph_i\cup\aleph_{j})$-LIL 
test (respectively, $(0.05,\aleph_i\cup\aleph_{i+1})$-LIL test).
\end{ex}

{\em Proof.} The probability could be calculated using either
equation (\ref{twoNvee}) or (\ref{twoNvee2}) in Theorem \ref{thmtwoNvee}
with $\theta=1-\alpha$. Our analysis shows that results 
from (\ref{twoNvee}) and (\ref{twoNvee2}) have a difference 
smaller than $0.000000001$ which is negligible. The values
in Table \ref{weakLIL01} are computed using
the equation (\ref{twoNvee}) and then verified using the
equation  (\ref{twoNvee2}).
\hfill{$\Box$}

\section{Weak-LIL test design II}
\label{weakdesignsec2}
In this section, we consider the design of weak $(\alpha, \aleph)$-LIL test
with $\aleph$ consisting at least three points.
To be consistent with Section \ref{weakdesignsec}, we use 
the following notations:
$\aleph_0=\{2^{0}n_1\}, \cdots$, and  $\aleph_t=\{2^{t}n_1\}$
for given $n_1$ and $t$. In particular, we will
consider the cases for $n_1=2^{26}$. 

\begin{theorem}
\label{thmthreeNvee}
For fixed $0<\alpha<1$ and $t_1,t_2\ge 2$, let $\theta=1-\alpha$,
$\aleph=\{n, t_1n, t_1t_2n\}$, and 
$\aleph_a=\{n,t_1n\}$. Then we have
\begin{equation}
\label{threeNvee}
\begingroup
\def\arraystretch{1.4} 
\begin{array}{l}
{\bf P}_{(\alpha, \aleph)}\simeq
{\bf P}_{(\alpha, \aleph_a)} + \\ \quad
\displaystyle\frac{1}{2\pi\sqrt{2\pi(t_1-1)}}
\displaystyle\int_{C_1}\displaystyle\int_{C_2}\displaystyle\int_{C_3}
e^{-\frac{x^2+y^2}{2}-\frac{(z-y)^2}{2(t_1-1)}}dxdydz
\end{array}
\endgroup
\end{equation}
where
$$
\begingroup
\def\arraystretch{2}
\begin{array}{l}
C_1=\left[-\theta\sqrt{2t_1\ln\ln t_1n}, \theta\sqrt{2t_1\ln\ln t_1n}\right]\\
C_2=\left[-\theta\sqrt{2\ln\ln n},\theta\sqrt{2\ln\ln n}\right]\\
C_3=\left[\sqrt{\frac{1}{t_2-1}}(\theta\sqrt{2t_2\ln\ln t_2t_1n}-z/\sqrt{t_1}),\infty\right).
\end{array}
\endgroup
$$
\end{theorem}

\noindent
{\em Proof.} By symmetry, it suffices to show that 
\begin{equation}
\label{eq3ninc}
\begingroup
\def\arraystretch{2}
\begin{array}{l}
Prob\left[S_{lil}(\xi\uzvert t_1t_2n)\ge \theta 
|\overline{{\bf E}_{\alpha, \aleph_a}}\right] \\ \quad
\simeq
\displaystyle\frac{1}{2\pi\sqrt{2\pi(t_1-1)}}
\displaystyle\int_{C_1}\displaystyle\int_{C_2}\displaystyle\int_{C_3}
e^{-\frac{x^2+y^2}{2}-\frac{(z-y)^2}{2(t_1-1)}}dxdydz
\end{array}
\endgroup
\end{equation}

By Corollary \ref{corn2nprob}, the probability that 
$S_{lil}(\xi\uzvert t_1n)\in [z,z+\Delta z]$ 
and $S_{lil}(\xi\uzvert t_1t_2n)>\theta$
is approximately
\begin{equation}
\label{initiiiii3}
P_{(z,\Delta z, t_1n)}\cdot
\int^{\infty}_{\sqrt{\frac{2}{t_2-1}}(\theta\sqrt{t_2\ln\ln t_2t_1n}-z\sqrt{\ln\ln t_1n})}\varphi(x)dx
\end{equation}
where 
$P_{(z,\Delta z, t_1n)}$ is the probability that 
$S_{lil}(\xi\uzvert t_1n)\in [z,z+\Delta z]$.
Let $\Delta_1=\sqrt{2t_1\ln\ln t_1n}\cdot \Delta z$.
By equation (\ref{probOF2n}) in the proof 
of Theorem \ref{thmtwoNvee}, the probability  $P_{(z,\Delta z, t_1n)}$
under the conditional event ``$\left|S_{lil}(\xi\uzvert n)\right|< \theta$''
is approximately 
\begin{equation}
\label{initi3t}
\begingroup
\def\arraystretch{2.5}
\begin{array}{l}
P_{(z,\Delta z, t_1n)}\simeq \\ 
\quad\displaystyle\frac{1}{2\pi}\displaystyle\int_{C_2}
\displaystyle\int_{\frac{z\sqrt{2t_1\ln\ln t_1n}-y}{\sqrt{t_1-1}}}^{\frac{z\sqrt{2t_1\ln\ln t_1n}+\Delta_1-y}{\sqrt{t_1-1}}} 
e^{-\frac{x^2+y^2}{2}} dxdy\\
\quad\simeq \displaystyle\int_{C_2}
\varphi(y) \varphi\left(\frac{z\sqrt{2t_1\ln\ln t_1n}-y}{\sqrt{t_1-1}}\right)
\frac{\Delta_1}{\sqrt{t_1-1}} dy\\
\quad\simeq \displaystyle\frac{\Delta_1}{\sqrt{t_1-1}} \displaystyle\int_{C_2}
\varphi(y)\cdot\varphi\left(\frac{z\sqrt{2t_1\ln\ln t_1n}-y}{\sqrt{t_1-1}}\right)dy
\end{array}
\endgroup
\end{equation}

By substituting (\ref{initi3t}) into (\ref{initiiiii3}), replacing
$z\sqrt{2t_1\ln\ln t_1n}$ with $w$, and  
integrating the obtained equation (\ref{initi3t}) over the interval
$w\in [-\theta\sqrt{2t_1\ln\ln t_1n}, \theta\sqrt{2t_1\ln\ln t_1n}]$,
and finally replacing the variable $w$ back to $z$, 
equation (\ref{eq3ninc}) is obtained. The theorem is then proved.
\hfill{$\Box$}

\begin{ex}
\label{3pointex}
Let $n_1=2^{26}$. By equation (\ref{threeNvee}) 
in Theorem \ref{thmthreeNvee}, we can calculate 
the following probabilities:
\begin{enumerate}
\item ${\bf P}_{(0.1, \aleph_0\cup\aleph_3\cup\aleph_6)}=0.07755$;
\item ${\bf P}_{(0.1, \aleph_0\cup\aleph_3\cup\aleph_8)}=0.07741$;
\item ${\bf P}_{(0.1, \aleph_0\cup\aleph_6\cup\aleph_8)}=0.07417$;
\item ${\bf P}_{(0.1, \aleph_3\cup\aleph_6\cup\aleph_8)}=0.06995$;
\item ${\bf P}_{(0.05, \aleph_0\cup\aleph_4\cup\aleph_8)}=0.05645$;
\end{enumerate}
By trying all different combinations, it can be shown that for any 
$\aleph=\aleph_{i_1}\cup \aleph_{i_2}\cup\aleph_{i_3}$ with different
$0\le i_1, i_2, i_3\le 8$, we have  $0.069\le {\bf P}_{0.1, \aleph}\le 0.08$
and $0.05\le {\bf P}_{0.05, \aleph}\le 0.06$.
\end{ex}

Theorem  \ref{thmthreeNvee} provides an algorithm for computing the 
probability ${\bf P}_{\alpha, \aleph}$ when $\aleph$  contains three points.
By recursively applying Corollary \ref{corn2nprob} as in the proof of 
Theorem  \ref{thmthreeNvee}, we can obtain algorithms for
calculating the probability ${\bf P}_{\alpha, \aleph}$ when 
$\aleph$ contains more than three points. 
The process is straightforward though tedious
and the details are omitted here. In the following,
we give an alternative approach to approximate the 
probability ${\bf P}_{(\alpha, \aleph)}$ with $|\aleph|>3$
by using Theorems \ref{thmtwoNvee} and \ref{thmthreeNvee}.

We show the approximation technique with the example 
of $\alpha=0.1$ and 
$\aleph=\aleph_0\cup \aleph_3\cup\aleph_6\cup\aleph_8$.
First we note that 
\begin{equation}
\label{eqa2}
\begin{array}{lll}
{\bf P}_{(\alpha,\aleph)} & =& 
{\bf P}_{(\alpha, \aleph_0\cup \aleph_3\cup\aleph_6)} +{\bf P}_{(\alpha,\aleph_8)}\\
&& 
-Prob\left[{\bf E}_{(\alpha,\aleph_8)}\cap
    {\bf E}_{(\alpha, \aleph_0\cup \aleph_3\cup\aleph_6)} \right]
\end{array}
\end{equation}
Since 
$$\begin{array}{l}
{\bf E}_{(\alpha,\aleph_8)}\cap
    {\bf E}_{(\alpha, \aleph_0\cup \aleph_3\cup\aleph_6)}=
({\bf E}_{(\alpha,\aleph_8)}\cap{\bf E}_{(\alpha,\aleph_0)})\\ \quad\quad\quad
  \cup  ({\bf E}_{(\alpha,\aleph_8)}\cap{\bf E}_{(\alpha,\aleph_3)})\cup
 ({\bf E}_{(\alpha,\aleph_8)}\cap{\bf E}_{(\alpha,\aleph_6)})
\end{array}$$
we have 
\begin{equation}
\label{eqa1}
\begingroup
\def\arraystretch{1.4}
\begin{array}{l} 
Prob\left[{\bf E}_{(\alpha,\aleph_8)}\cap
    {\bf E}_{(\alpha, \aleph_0\cup \aleph_3\cup\aleph_6)} \right]\\
=Prob\left[{\bf E}_{(\alpha,\aleph_0)}\cap
{\bf E}_{(\alpha,\aleph_8)}\right]+
Prob\left[{\bf E}_{(\alpha,\aleph_3)}\cap
{\bf E}_{(\alpha,\aleph_8)}\right]\\
\quad +Prob\left[{\bf E}_{(\alpha,\aleph_6)}\cap
{\bf E}_{(\alpha,\aleph_8)}\right]\\
\quad - Prob\left[{\bf E}_{(\alpha,\aleph_0)}\cap{\bf E}_{(\alpha,\aleph_3)}\cap
{\bf E}_{(\alpha,\aleph_8)}\right]\\
\quad - Prob\left[{\bf E}_{(\alpha,\aleph_0)}\cap{\bf E}_{(\alpha,\aleph_6)}\cap
{\bf E}_{(\alpha,\aleph_8)}\right]\\
\quad - Prob\left[{\bf E}_{(\alpha,\aleph_3)}\cap{\bf E}_{(\alpha,\aleph_6)}\cap
{\bf E}_{(\alpha,\aleph_8)}\right]\\
\quad +2\cdot 
 Prob\left[{\bf E}_{(\alpha,\aleph_0)}\cap {\bf E}_{(\alpha,\aleph_3)}\cap{\bf E}_{(\alpha,\aleph_6)}\cap
{\bf E}_{(\alpha,\aleph_8)}\right]\\
\end{array}
\endgroup
\end{equation}
Let $\varepsilon=Prob\left[{\bf E}_{(\alpha,\aleph_0)}\cap {\bf E}_{(\alpha,\aleph_3)}\cap{\bf E}_{(\alpha,\aleph_6)}\cap
{\bf E}_{(\alpha,\aleph_8)}\right]$.
By substituting (\ref{eqa1}) into (\ref{eqa2}) and simplifying it,
we get 
\begin{equation}
\label{eqa3}
\begingroup
\def\arraystretch{1.4}
\begin{array}{lll}
{\bf P}_{(\alpha,\aleph)} & =& 
\sum_{i\in \{0,3,6,8\}}{\bf P}_{(\alpha,\aleph_i)}\\
&&+\sum_{i_1,i_2,i_3\in\{0,3,6,8\}} {\bf P}_{(\alpha,\aleph_{i_1}\cup\aleph_{i_2}\cup\aleph_{i_3} )}\\
&&-\sum_{i_1,i_2\in \{0,3,6,8\}}{\bf P}_{(\alpha,\aleph_{i_1}\cup\aleph_{i_2})}-2\varepsilon\\
&\simeq &0.09662-2\varepsilon
\end{array}
\endgroup
\end{equation}
On the other hand, we have 
$$
\begingroup
\def\arraystretch{1.4}
\begin{array}{lll}
2\varepsilon &<& 2\cdot Prob\left[{\bf E}_{(\alpha,\aleph_3)}
  \cap{\bf E}_{(\alpha,\aleph_6)}\cap{\bf E}_{(\alpha,\aleph_8)}\right]\\
&=&{\bf P}_{(\alpha,\aleph_{3}\cup\aleph_{6}\cup\aleph_{8} )}+
\sum_{i\in \{3,6,8\}}{\bf P}_{(\alpha,\aleph_i)}\\
&& -\sum_{i_1,i_2\in \{3,6,8\}}{\bf P}_{(\alpha,\aleph_{i_1}\cup\aleph_{i_2})}\\
&\simeq& 0.00032
\end{array}
\endgroup
$$
Thus we have 
$0.09630<{\bf P}_{(\alpha,\aleph)}<0.09662.$
In other words, a random sequence passes
the weak $(0.1, \aleph_0\cup \aleph_3\cup\aleph_6\cup\aleph_8)$-LIL
test with approximately $9.65\%$ probability.

\section{Strong LIL test design}
\label{strongLILdesign}
This section considers the following strong LIL tests.

\noindent
{\bf Strong LIL Test}:
Let $\alpha\in (0, 0.25]$ and $\aleph_a,\aleph_b,\aleph_c\subset N$
be subsets of natural numbers.
We say that a sequence $\xi$  passes the strong
$(\alpha;\aleph_a,\aleph_b$)-LIL test 
if there exist $n_1\in\aleph_a$ and $n_2\in\aleph_b$
such that 
\begin{equation}
\label{strongLILformula}
\begingroup
\def\arraystretch{1.4}
\begin{array}{l}
\left|S_{lil}(\xi\uzvert n_i)\right|>1-\alpha \mbox{ for }i=1,2;\\
S_{lil}(\xi\uzvert n_1)S_{lil}(\xi\uzvert n_2)<0.
\end{array}
\endgroup
\end{equation}
Alternatively, we say that a sequence $\xi$  passes the strong
$(\alpha;\aleph_c$)-LIL test 
if there exist $n_1,n_2\in\aleph_c$
such that (\ref{strongLILformula}) holds.
Furthermore, ${\bf SP}_{(\alpha; \aleph_a,\aleph_b)}$ and  
${\bf SP}_{(\alpha; \aleph_c)}$
denote the probability that a random sequence passes the strong
$(\alpha;\aleph_a,\aleph_b$)-LIL and $(\alpha;\aleph_c$)-LIL
tests respectively.

\begin{theorem}
\label{strongthm}
For fixed $0<\alpha<1$ and $t\ge 2$, let
$\theta=1-\alpha$, $\aleph_a=\{n\}$, and 
 $\aleph_b=\{tn\}$. We have
\begin{equation}
\label{StrongLILNvee2} 
\begingroup
\def\arraystretch{1.4}
\begin{array}{l}
{\bf SP}_{(\alpha, \aleph_a,\aleph_b)}\simeq \\ \quad
\displaystyle\frac{1}{\pi}\displaystyle\int_{\theta\sqrt{2\ln\ln n}}^{\infty}
\displaystyle\int_{-\infty}^{-\sqrt{\frac{1}{t-1}}(\theta\sqrt{2t\ln\ln tn}+y)} e^{-\frac{x^2+y^2}{2}} dxdy
\end{array}
\endgroup
\end{equation}
\end{theorem}

\noindent
{\em Proof.}
The theorem could be proved in a similar way as in the proof of Theorem 
\ref{thmtwoNvee}.
\hfill{$\Box$}

\begin{ex}
\label{slilex}
Let $\alpha=0.1$, $\aleph_0=\{2^{26}\}$, $\aleph_7=\{2^{33}\}$, 
and $\aleph_8=\{2^{34}\}$. Then we have  
${\bf SP}_{(\alpha, \aleph_0,\aleph_7)}\simeq 0.0001981$ and 
${\bf SP}_{(\alpha, \aleph_0,\aleph_8)}\simeq 0.0002335$
\end{ex}

In the following, we provide another approach for 
obtaining better probability bounds for strong LIL tests.
In a negative binomial distribution $f(k;r,\frac{1}{2})$
denote the probability that the $r$th one appears at the position
$r+k$. It is well known that for this distribution, we have mean
$\mu=r$ and variance $\sigma=\sqrt{2r}$. Thus the probability 
that the $r$'s one appears before the $n$th position is approximated
by the following probability:
\begin{equation}
\label{negaBf}
\frac{1}{2\sqrt{r\pi}}\int_{-\infty}^n e^{-\frac{(x-2r)^2}{4r}}dx
\end{equation}
For $n_1=2^{26}$ and $n_2=2^{34}$, assume that 
$S_{lil}(\xi\uzvert n_1)\le -y$ for given $y\ge \theta$. Then we have
\begin{equation}
\label{n1portion1} 
S(\xi\uzvert n_1)\le\frac{n_1-y\sqrt{2n_1\ln \ln n_1}}{2}
\end{equation}
By (\ref{n1portion1}), in order for $S_{lil}(\xi\uzvert n_2)\ge \theta$, 
we need to have 
\begin{dmath}
r(y)=S(\xi[n_1..n_2-1])\ge \frac{n_2+\theta\sqrt{2n_2\ln\ln n_2}
-n_1+y\sqrt{2n_1\ln\ln n_1}}{2}
\end{dmath}
Let $\alpha=1-\theta$,  $\aleph_a=\{n_1\}$, and 
$\aleph_b=\{n_2\}$.
Using the same argument as in the proof of Theorem \ref{thmtwoNvee}
(in particular, the arguments for integrating equation (\ref{initiiiii}))
and the negative binomial distribution equation (\ref{negaBf}), 
the probability  that a sequence passes the strong 
$(\alpha; \aleph_a, \aleph_b)$-LIL test can be calculated with the 
following equation.
\begin{equation}
\label{slileqbn}
\frac{1}{\pi}\int_{-\infty}^{-\theta\sqrt{2\ln\ln n_1}}
 \int_{-\infty}^{n_2-n_1}\frac{1}{\sqrt{2r(y)}} 
e^{-\frac{y^2}{2}-\frac{(x-2r(y))^2}{4r(y)}}dxdy
\end{equation}
By substituting the values of $\theta$, $n_1$, and $n_2$,
(\ref{slileqbn}) evaluates to 0.0002335.
In other words, a random sequence passes the strong 
$(0.1;\aleph_0,\aleph_8)$-LIL test with probability $0.023\%$
(this value is same as the value in Example \ref{slilex}).

Both (\ref{StrongLILNvee2}) and (\ref{slileqbn})
could be used to calculate the probability for strong LIL tests.
These equations could be used to generalize results in Example \ref{slilex}
to cases of strong $(\alpha; \aleph_a,\aleph_b)$-LIL
test with multiple points in $\aleph_b$.

\section{Evaluating Pseudorandom Generators}
\label{evaRsec}
In order to evaluate the quality of a pseudorandom generator ${\cal G}$,
we first choose a fixed $n$ of sequence length, 
a value $0<\alpha\le 0.1$, and mutually 
distinct subsets $\aleph_0, \cdots, \aleph_t$ of $\{1,\cdots, n\}$. 
It is preferred that the $S_{lil}$ values on these subsets are 
as independent as possible (though they are impossible to be independent). 
For example, we may choose 
$\aleph_i$ as in Section \ref{weakdesignsec2}. Then we can carry out the 
following steps.
\begin{enumerate}
\item Set ${\bf P}_{(\alpha, \aleph)}^+={\bf P}_{(\alpha, \aleph)}^-=
\frac{1}{2}{\bf P}_{(\alpha, \aleph)}$ for all $\aleph$.
\item Use ${\cal G}$ to construct a set of $m\ge 100$ binary sequences of 
length $n$.
\item For each $\aleph$, calculate probability
$P_{(\alpha,\aleph)}^+$ that these sequences pass the weak 
$(\alpha,\aleph_i)$-LIL test via $S_{lil}\ge 1-\alpha$
(respectively, $P_{(\alpha,\aleph)}^-$ for $S_{lil}\le -1+\alpha$).
\item Calculate the average absolute probability distance 
\begin{dmath*}
\Delta_{wlil}= \frac{1}{t+1}\sum_{i=0}^{t}{\bf P}_{(\alpha, \aleph_i)}^{-1}\left(\left|P_{(\alpha,\aleph_i)}^+-{\bf P}_{(\alpha,\aleph_i)}^+\right|+
\left|P_{(\alpha,\aleph_i)}^--{\bf P}_{(\alpha,\aleph_i)}^-\right|\right)
\end{dmath*}
and the root-mean-square deviation 
\begin{dmath*}
\label{lilrmsd}
\mbox{RMSD}_{wlil}=\sqrt{\frac{\sum_{0\le i\le j\le t}
\left(p_{i,j,1}^2+p_{i,j,2}^2\right)}{(t+1)(t+2)}}
\end{dmath*} 
where $p_{i,j,1}^+=P_{(\alpha,\aleph_i\cup\aleph_j)}^+-{\bf P}_{(\alpha,\aleph_i\cup\aleph_j)}^+$
and $p_{i,j,2}^+=P_{(\alpha,\aleph_i\cup\aleph_j)}^--{\bf P}_{(\alpha,\aleph_i\cup\aleph_j)}^-$
\item Decision criteria: the smaller $\Delta_{wlil}$ and 
$\mbox{RMSD}_{wlil}$, the better generator ${\cal G}$.
\end{enumerate}

\section{Snapshot LIL tests and random generator evaluation}
\label{snapshotsec}
We have considered statistical tests based on the limit theorem
of the law of the iterated logarithm. These tests do not take 
full advantage of the distribution $S_{lil}$, which
defines a probability measure on the real line $R$.
Let ${\cal R}\subset \Sigma^n$ be a set of $m$ sequences
with a standard probability definition on it. That is,
for each $x_0\in{\cal R}$, let $Prob[x=x_0]=\frac{1}{m}$.
Then each set ${\cal R}\subset \Sigma^n$ 
induces a probability measure $\mu_n^{\cal R}$ on $R$ 
by letting 
$$\mu_n^{\cal R}\left(I\right)=Prob\left[S_{lil}(x)\in I, x\in{\cal R}\right]$$
for each Lebesgue measurable set $I$ on $R$.
For $U= \Sigma^n$, we use $\mu_n^U$ to denote the
corresponding probability measure induced 
by the uniform distribution. By Definition
\ref{defnmartingaleind}, if ${\cal R}_n$ is the 
collection of all length $n$ sequences generated by a pseudorandom generator,
then the difference between $\mu_n^U$ and 
$\mu_n^{{\cal R}_n}$ is negligible.

By Theorem \ref{approximation} 
and Lemma \ref{dlaplaceslil}, for a uniformly
chosen $\xi$, the distribution
of $S^*(\xi\uzvert n)$ could be approximated by a normal distribution
of mean $0$ and variance $1$, with error bounded by 
$\frac{1}{n}$ (see \cite{feller}). In other words,
the  measure  $\mu_n^U$ can be calculated as
\begin{dmath}
\mu_n^U((-\infty,x])\simeq \Phi(x\sqrt{2\ln\ln n})
=\sqrt{2\ln\ln n}\int_{-\infty}^x\phi(y\sqrt{2\ln\ln n})dy.
\end{dmath}
Table \ref{inducedProbslil} 
in the Appendix lists values  $\mu_n^U(I)$ for $0.05$-length 
intervals $I$ with $n=2^{26}, \cdots, 2^{34}$. 

In order to evaluate a pseudorandom generator $G$, first 
choose a sequence of testing points $n_0, \cdots, n_t$ (e.g.,
$n_0=2^{26+t}$). Secondly
use $G$ to generate a set ${\cal R}\subseteq \Sigma^{n_t}$ of $m$ 
sequences.  Lastly compare the distances between 
the two probability measures $\mu_n^{\cal R}$ and  $\mu_n^U$
for $n=n_0,\cdots, n_t$.  

A generator $G$ is considered ``good'',
if for sufficiently large $m$, the distances between 
$\mu_n^{\cal R}$ and  $\mu_n^U$ are negligible (or smaller than a given 
threshold). There are various  definitions of statistical
distances for probability measures. In our analysis, we will consider
the total variation distance \cite{clarkson1933definitions}
\begin{equation}
\label{totaldistancefor}
d(\mu_n^{\cal R},\mu_n^U)=\sup_{A\subseteq{\cal B}}\left| \mu_n^{\cal R}(A)-\mu_n^U(A)\right|
\end{equation}
Hellinger distance \cite{hellinger1909neue}
\begin{equation}
\label{divergenceformula}
H(\mu_n^{\cal R}||\mu_n^U)=\frac{1}{\sqrt{2}}
\sqrt{
\sum_{A\in {\cal B}}\left(\sqrt{\mu_n^{\cal R}(A)}-\sqrt{\mu_n^U(A)}\right)^2
}
\end{equation}
and the root-mean-square deviation
\begin{dmath}
\mbox{RMSD}(\mu_n^{\cal R},\mu_n^U)
=\sqrt{\frac{\sum_{A\in {\cal B}}\left({\mu_n^{\cal R}(A)}-{\mu_n^U(A)}\right)^2
}{42}
}
\end{dmath}
where  ${\cal B}$ is a partition of the real line $R$ that is defined as 
$$\{(\infty, 1),[1,\infty)\}\cup
\left\{[0.05x-1,0.05x-0.95): 0\le x\le  39\right\}.$$

In Section \ref{experisection}, we will present some examples 
of using  these distance to evaluate several pseudorandom generators.

\section{Experimental results}
\label{experisection}
As an example to illustrate the importance of LIL tests, we carry out 
weak LIL test experiments on pseudorandom
generators SHA1PRNG (Java) and NIST DRBG \cite{nist80022}
with parameters $\alpha = 0.1$ (and $0.05$) and
$\aleph_0=\{2^{26}\}, \cdots, \aleph_8=\{2^{34}\}$ 
(note that $2^{26}$bits$=8$MB and $2^{34}$bits$=2$GB).

Before carrying out LIL based tests, we run 
NIST SP800-22 test tool \cite{nistrngtool}
on sequences that have been generated.
The test tool  \cite{nistrngtool} only checks
the first 1,215,752,192 bits ($\simeq$145MB) 
of a sequences since the software 
uses 4-byte {\tt int} data type for integer variables. 
The initial 145MB of each sequence that we have generated 
passes  NIST tests with P-values larger than $0.01$ 
except for the ``longest run of ones in a block'' 
test which failed for several sequences.

\subsection{Java SHA1PRNG API based sequences}
\label{sunrandomsec}
\begin{figure}[htb]
\caption{LIL test results for sequences generated by Java SHA1PRNG, NIST SP800 90A SHA1-DRBG, and
NIST SP800 90A SHA2-DRBG}
\label{nistsha1all}
\begin{center}
\includegraphics[width=0.45\textwidth]{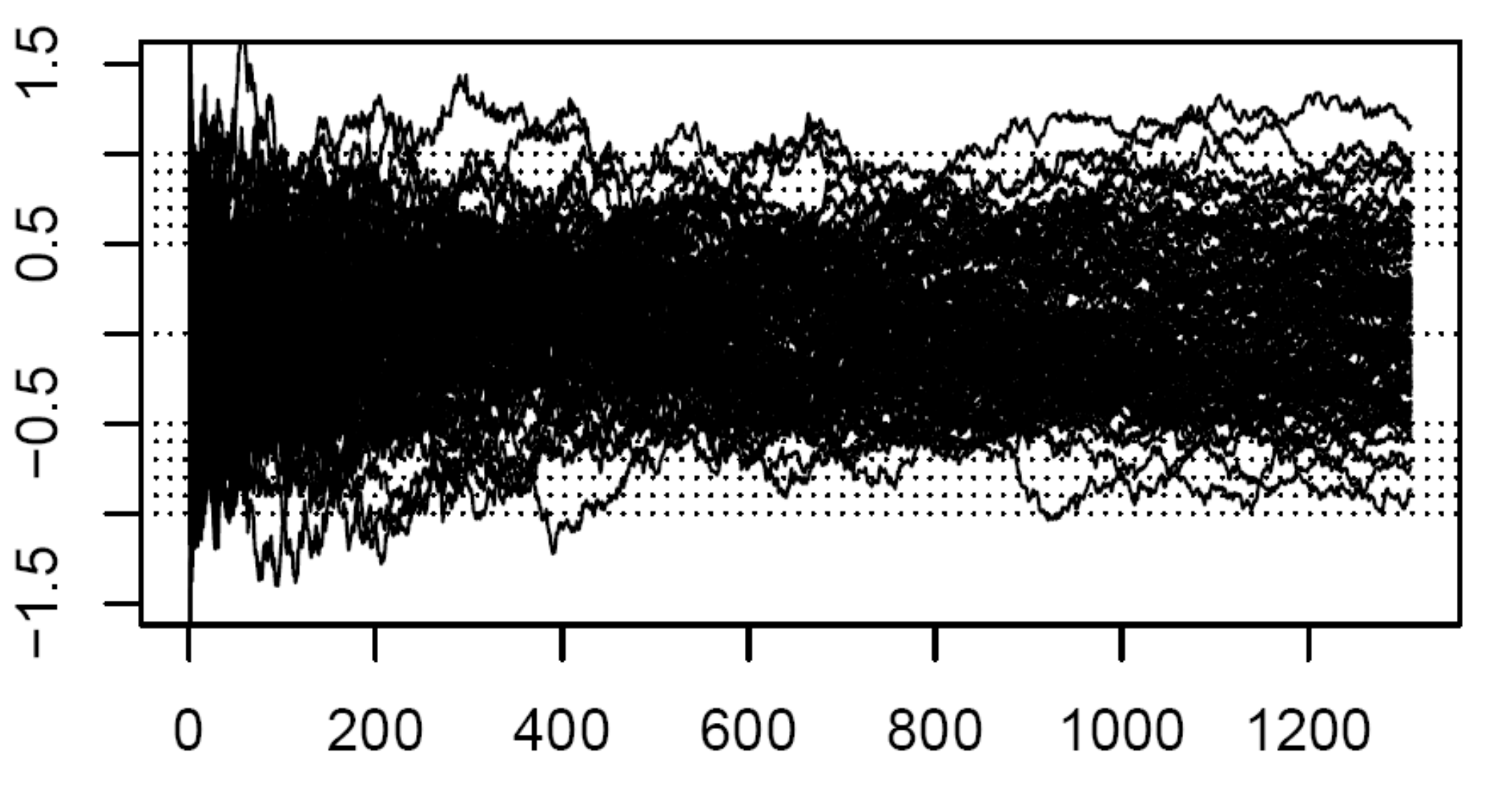} 
\includegraphics[width=0.45\textwidth]{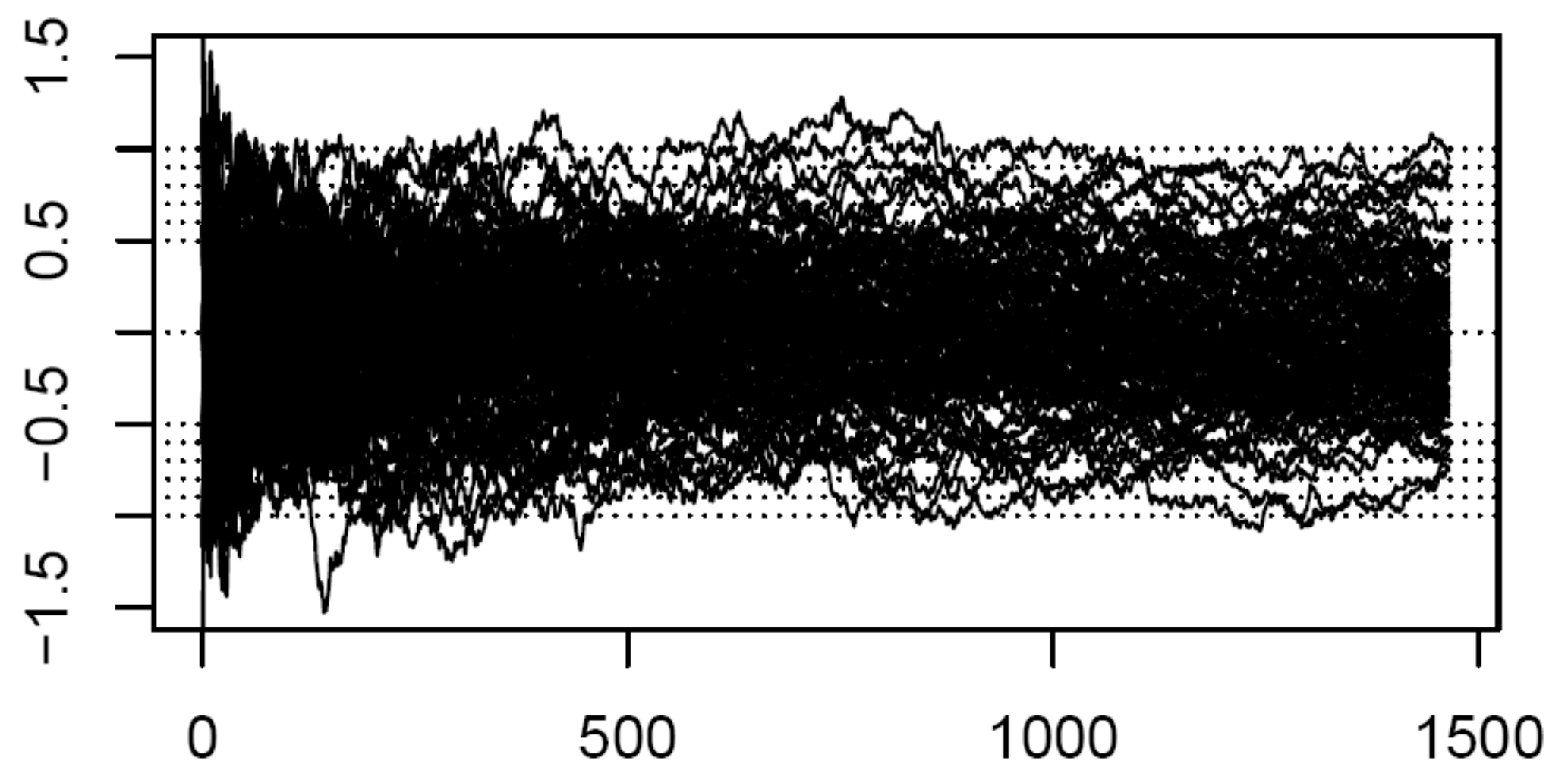} 
\includegraphics[width=0.45\textwidth]{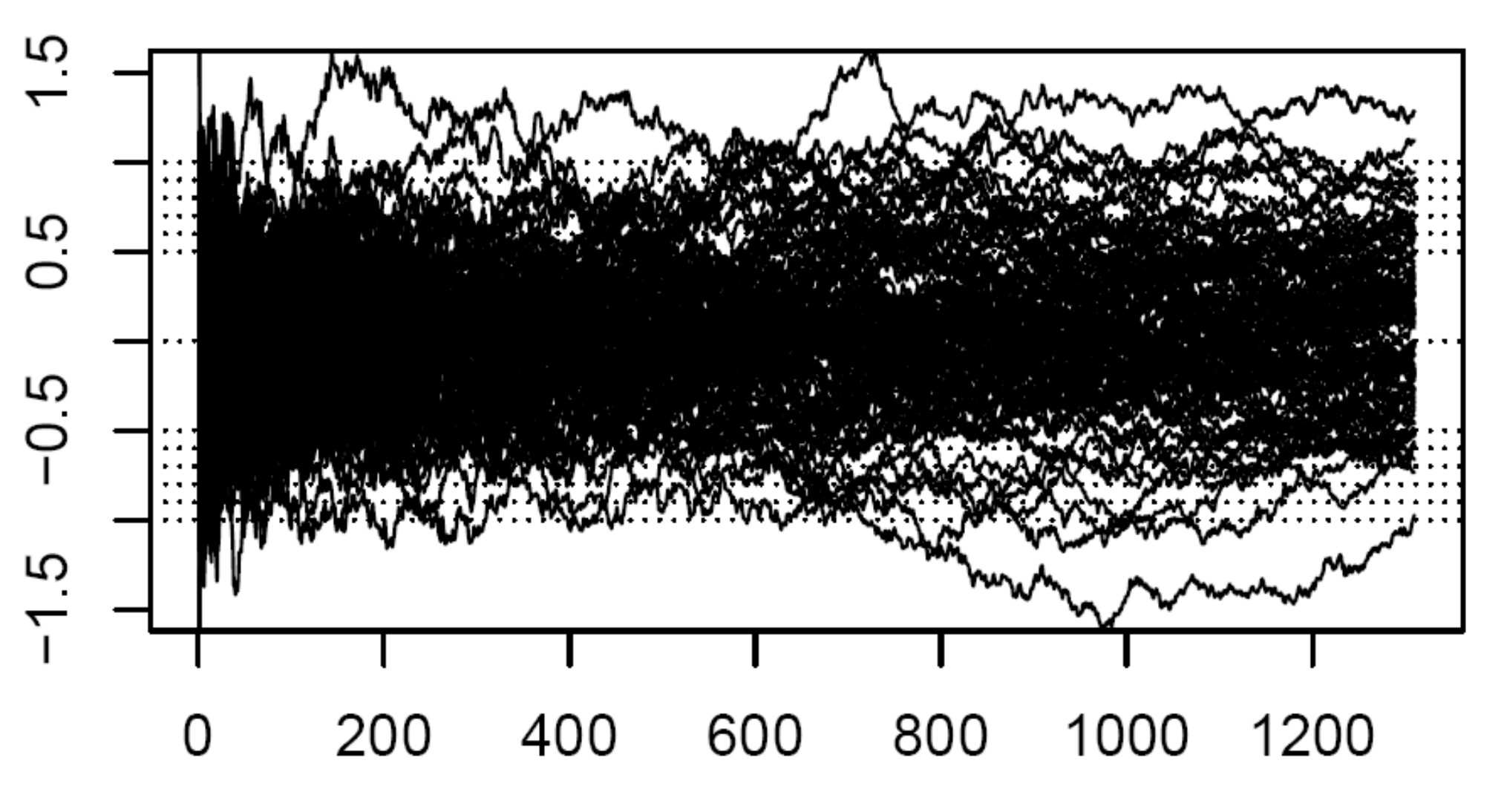} 
\end{center}
\end{figure} 
The pseudorandom generator SHA1PRNG API in Java 
generates sequences
$\mbox{SHA1}'(s, 0)\mbox{SHA1}'(s,1)\cdots$,
where $s$ is an optional seeding string of arbitrary length,  
the counter $i$ is 64 bits long, and 
$\mbox{SHA1}'(s, i)$ is the first 64 bits of
$\mbox{SHA1}(s, i)$.
In the experiment, we generated one thousand of sequences 
with four-byte seeds of integers $0,1, 2, \cdots, 999$ respectively.
For each sequence generation,
the ``random.nextBytes()'' method of  SecureRandom Class
is called $2^{26}$ times and a $32$-byte output is requested for each call.
This produces sequences of $2^{34}$ bits long. 
The LIL test is then run on these sequences and 
the first picture in Figure \ref{nistsha1all} 
shows the LIL-test result curves for the first 100 sequences.
To reduce the size of the figure, 
we use the scale $10000n^2$ for the $x$-axis. In other words,
Figure \ref{nistsha1all}  shows the values 
$S_{lil}(\xi[0..10000n^2-1])$ for $1\le n\le 1310$.
At this scale, the points $\aleph_0,\cdots, \aleph_8$
are mapped to 82, 116, 164, 232, 328, 463, 655, 927, and 1310
respectively.  Table \ref{scalemap} shows the number of sequences
that reach the value 0.9, -0.9, 0.95, and -0.95 at 
corresponding testing points respectively.

\begin{table}[ht]
\caption{Number of sequences that pass the LIL values $0.9$ and $0.95$}
\label{scalemap}
\begin{center}
{
\scriptsize
\begin{tabular}{|c|c|c|c|c|c|c|c|c|c|}\hline
$\aleph$ &$\aleph_0$ & $\aleph_1$ & $\aleph_2$ & $\aleph_3$ & $\aleph_4$ & 
    $\aleph_5$& $\aleph_6$ & $\aleph_7$ & $\aleph_8$ \\ \hline
$n$& 82&116&164&232&328&463&655&927&1310\\ \hline
\multicolumn{10}{|c|}{Java SHA1PRNG}\\ \hline
$0.9$& 20&16&20&20&16&14&17&11&11\\ \hline
$-0.9$&18&20&18&17&14&11&12&11&9\\ \hline
$0.95$&14&12&13&18&12&10&15&7&8\\ \hline
$-0.95$&13&13&14&9&10&7&9&8&6\\ \hline
\multicolumn{10}{|c|}{NIST SP800 90A SHA1-DRBG at sample size 1000}\\ \hline
$0.9$  &15&16&15&12&8&9&17&10&8\\ \hline
$-0.9$ &15&19&12&18&10&16&14&9&`2\\ \hline
$0.95$ &10&9&12&10&5&5&11&6&6\\ \hline
$-0.95$&11&12&8&13&8&10&10&7&12\\ \hline
\multicolumn{10}{|c|}{NIST SP800 90A SHA256-DRBG  at sample size 1000}\\ \hline
$0.9$& 13&16&14&20&13&15&21&16&9\\ \hline
$-0.9$&16&13&14&5&13&9&11&13&10\\ \hline
$0.95$&9&10&12&15&9&10&16&14&3\\ \hline
$-0.95$&13&9&8&4&8&6&9&12&9\\ \hline
\multicolumn{10}{|c|}{NIST SP800 90A SHA256-DRBG  at sample size 10000}\\ \hline
$0.9$&164&157&162&145&128&128&133&121&114\\ \hline
$-0.9$&154&142&142&130&123&128&123&120&107\\ \hline
$0.95$&120&107&127&110&89&93&93&84&70\\ \hline
$-0.95$&107&106&92&99&91&93&95&84&78\\ \hline
\end{tabular}
}
\end{center}
\end{table}
Using the partition set ${\cal B}$, the probability 
distributions  $\mu_n^{JavaSHA1}$  
with $n=2^{26}, \cdots, 2^{34}$ are presented in the Appendix 
Table \ref{inducedProbieee}.
Figure \ref{snapShotJava} compares these distributions.
\begin{figure}[htb]
\caption{The distributions $\mu_n^U$ and  $\mu_n^{JavaSHA1}$  
with $n=2^{26}, \cdots, 2^{34}$}
\label{snapShotJava}
\begin{center}
\includegraphics[width=0.45\textwidth]{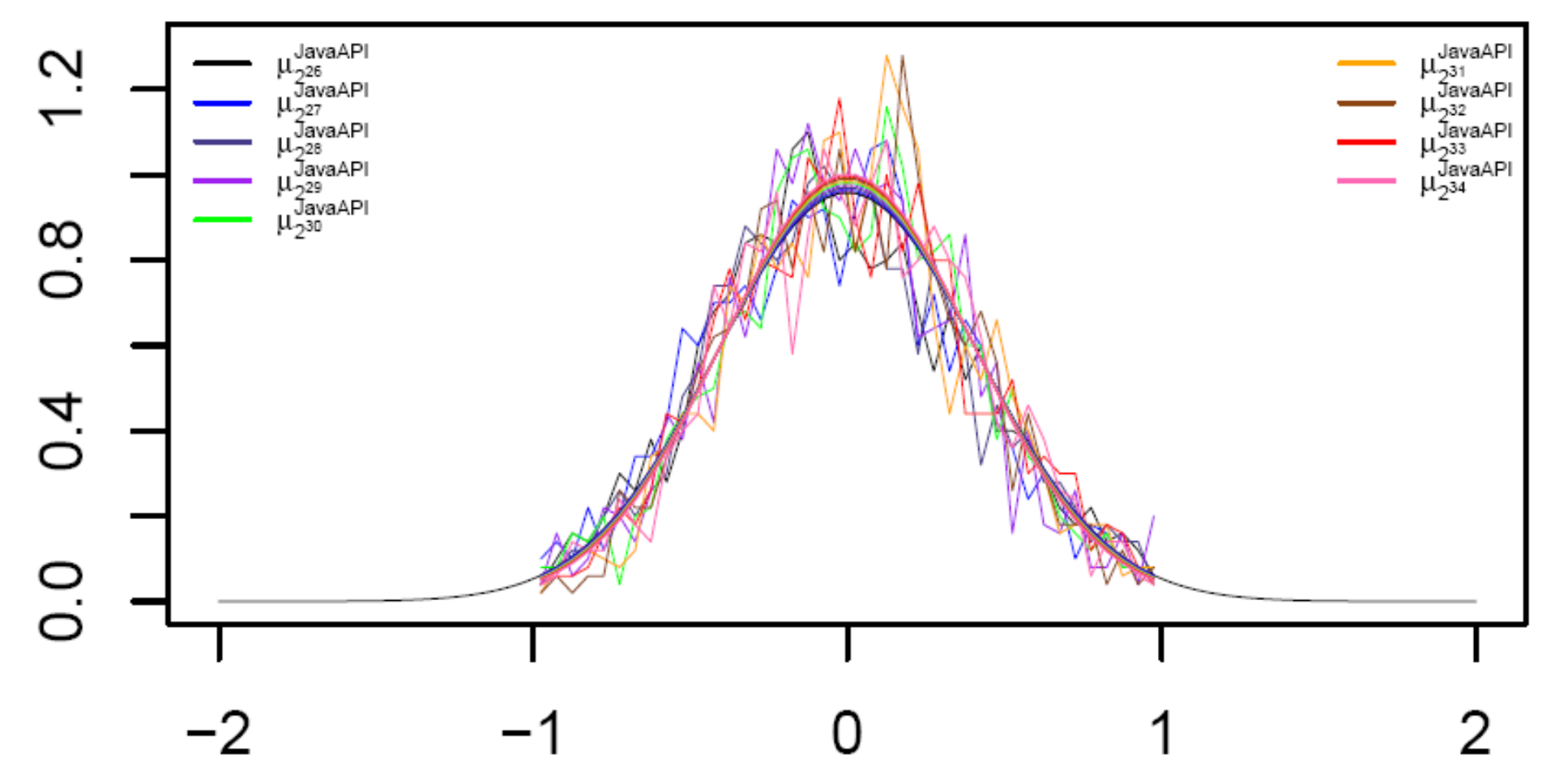} 
\end{center}
\end{figure}

\subsection{NIST SP800 90A DRBG pseudorandom generators}
\label{grandomsec}
NIST SP800.90A  \cite{nist80090a} specifies three 
types of DRBG generators:
hash function based, block cipher based, and ECC based.
For DRBG generators,  
the maximum number of calls between reseeding
is $2^{48}$ for hash function 
and AES based generators (the number is $2^{32}$ for 
T-DES and ECC-DRBG generators). 
In our experiment, we used hash function based DRBG.
where a hash function $G$ is used to generate sequences
$G(V)G(V+1)G(V+2)\cdots$ with $V$ being seedlen-bit counter
that is derived from the secret seeds. The seedlen is $440$ 
for SHA1 and SHA-256 and the value of $V$ is revised 
after at most $2^{19}$ bits are output.
We generated 10000 sequences nistSHADRBG0, $\cdots$, nistSHADRBG9999.
For each sequence nistSHADRBG$i$, the seed
``$i$th secret seed for NIST DRBG" is used to derive 
the initial DRBG state $V_0$ 
and $C_0$.  Each sequence is of the format 
$G(V_0)G(V_0)\cdots G(V_0+2^{12}-1)G(V_1)G(V_1+1)\cdots G(V_{2^{25}}+2^{12}-1)$,
where $V_{i+1}$ is derived from the value of $V_i$ and $C_i$.
In other words, each $V$ is used $2^{12}$ times before it is revised. 
The second picture (respectively, the third picture) 
in Figure \ref{nistsha1all} 
shows the LIL-test result curves for the first 100 sequences
when SHA1 (respectively, SHA256) is used as 
the hash function and  Table \ref{scalemap} 
shows the number of sequences
that reach the value 0.9, -0.9, 0.95, and -0.95 at corresponding 
testing points respectively.

\begin{figure}[htb]
\caption{The distributions $\mu_n^U$ and $\mu_n^{nistDRBGsha1}$  
with $n=2^{26}, \cdots, 2^{34}$}
\label{snapShotnistSHA1}
\begin{center}
\includegraphics[width=0.45\textwidth]{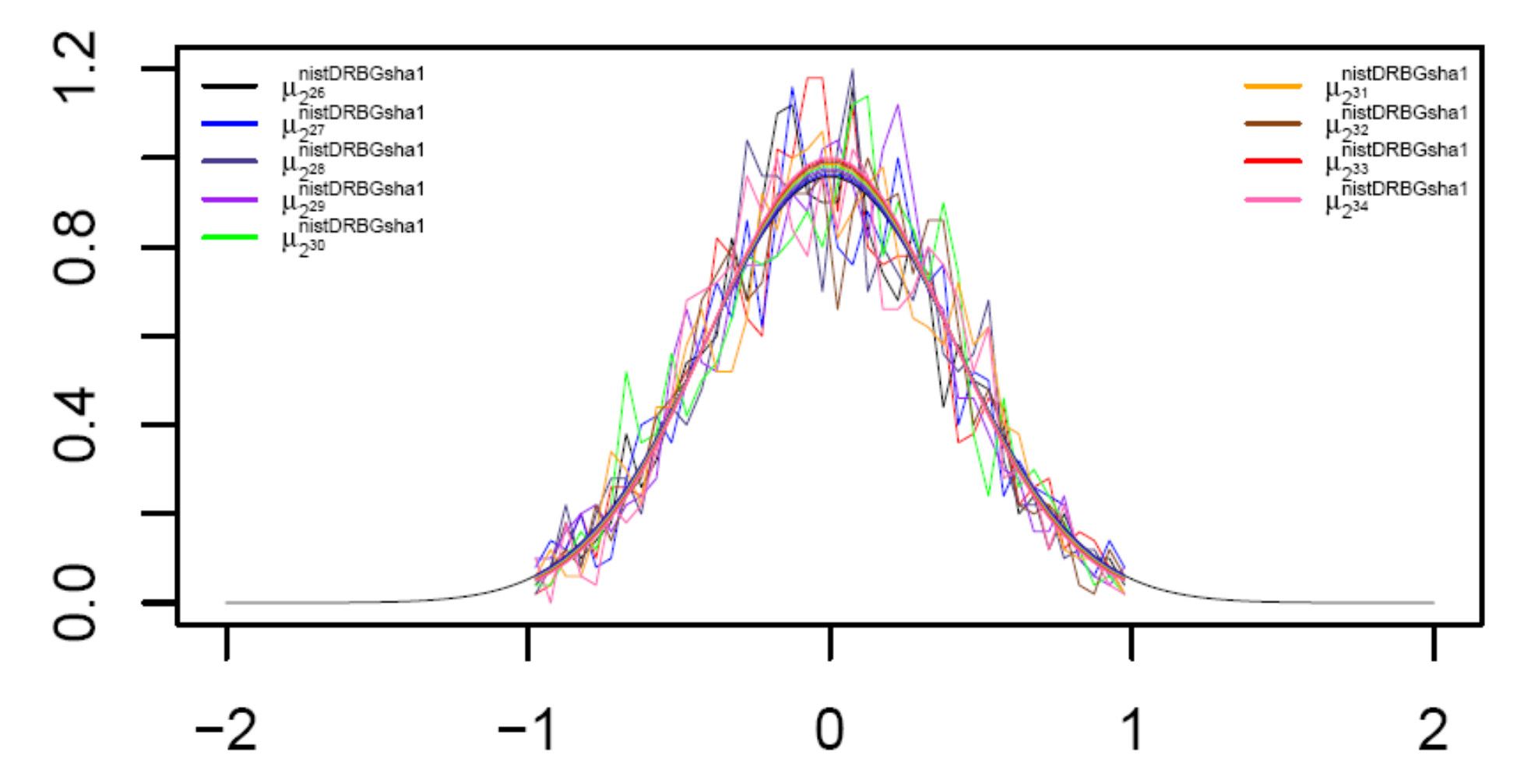} 
\end{center}
\end{figure}

The probability distributions  $\mu_n^{nistDRBGsha1,1000}$, 
$\mu_n^{nistDRBGsha256,1000}$, and $\mu_n^{nistDRBGsha256,10000}$ on partition ${\cal B}$
are presented in Appendix Tables \ref{inducedProbnistSHA1},
\ref{inducedProbnistSHA256}, \ref{inducedProbnistSHA25610k} respectively. 
Figure \ref{snapShotnistSHA1} compares the distributions
$\mu_n^{nistDRBGsha1}$ at the sample size 1000. The comparisons 
for  $\mu_n^{nistDRBGsha256}$ 
are presented in Appendix Figure \ref{snapShotnistSHA256}
and \ref{snapShotnistSHA25610k}.

\subsection{Comparison and Discussion}
Based on Table \ref{scalemap}, 
the average absolute probability distance $\Delta_{wlil}$ 
and the root-mean-square deviation $\mbox{RMSD}_{wlil}$ at the sample size 1000
(for DRBG-SHA256, we also include results for sample size 10000)
are calculated and shown in Table \ref{distresults}. These 
values are quite large.
\begin{table*}[ht]
\caption{The probability distances $\Delta_{wlil}$ and  $\mbox{RMSD}_{wlil}$}
\label{distresults}
\begin{center}
{
\scriptsize
\begin{tabular}{|c|c|c|c|c|}\hline
  &Java SHA1PRNG & DRBG-SHA1 & DRBG-SHA256 (1000)  & DRBG-SHA256 (10000) \\ \hline
$\Delta_{wlil, 0.1}$ &  0.140       &  0.194    & 0.200 &0.045\\ \hline 
$\Delta_{wlil, 0.05}$&  0.276       &  0.224     & 0.289&0.063\\ \hline
$\mbox{RMSD}_{wlil, 0.1}$&  0.004647&0.003741&0.004984&0.00118\\ \hline
$\mbox{RMSD}_{wlil, 0.05}$& 0.004042&0.003023&0.004423&0.001107\\ \hline
\end{tabular}
}
\end{center}
\end{table*}

Based on snapshot LIL tests at points
$2^{26}, \cdots, 2^{34}$, the corresponding 
total variation distance $d(\mu^{\cal R}_n,\mu_n^U)$, 
Hellinger distance $H(\mu^{\cal R}_n||\mu_n^U)$,
and the root-mean-square deviation
$\mbox{RMSD}(\mu_n^{\cal R},\mu_n^U)$ at sample size 1000
(also DRBG-SHA256 at sample size 10,000)
are calculated and shown in Table \ref{snaplilresulttable},
where subscripts $1,2,3,4$ are for JavaSHA1, nistDRBGsha1, 
nistDRBGsha256 (sample size 1000), and nistDRBGsha256 (sample size 10000) 
respectively. 
It is observed that at the sample size 1000,
the average distance between  $\mu^{\cal R}_n$ and 
$\mu_n^U$ is larger than 0.06 and 
the root-mean-square deviation is around 0.005.
\begin{table}[ht]
\caption{total variation and Hellinger distances}
\label{snaplilresulttable}
\begin{center}
{
\scriptsize
\begin{tabular}{|c|c|c|c|c|c|c|c|c|c|}\hline
$n$ &$2^{26}$&$2^{27}$&$2^{28}$&$2^{29}$&$2^{30}$&$2^{31}$&$2^{32}$& $2^{33}$&$2^{34}$\\ \hline
$d_1$&.074&.704&.064&.085&.067&.085&.074&.069&.071\\ \hline
$H_1$&.062&.067&.063&.089&.066&.078&.077&.061&.068\\ \hline
$\mbox{RMSD}_1$&.005&.005&.004&.005&.004&.006&.005&.005&.005\\ \hline
$d_2$&.066&.072&.079&.067&.084&.073&.065&.078&.083 \\ \hline
$H_2$&.060&.070&.073&.062&.077&.066&.067&.070&.087 \\ \hline
$\mbox{RMSD}_2$&.004&.005&.005&.004&.005&.004&.004&.005&.005\\ \hline
$d_3$&.076&.069&.072&.093&.071&.067&.078&.081&.066 \\ \hline
$H_3$&.082&.064&.068&.088&.079&.073&.076&.074&.080 \\ \hline
$\mbox{RMSD}_3$&.005&.004&.004&.006&.004&.004&.005&.005&.005\\ \hline
$d_4$&.021&.022&.026&.024&.022&.024&.026&.024&.021\\ \hline
$H_4$&.019&.021&.024&.024&.022&.023&.025&.022&.021 \\ \hline
$\mbox{RMSD}_4$&.001&.001&.002&.001&.001&.002&.002&.002&.001\\ \hline
\end{tabular}
}
\end{center}
\end{table}

Though the statistical distances in 
Tables \ref{distresults} and  \ref{snaplilresulttable} 
may be acceptable for various applications,
for a cryptographic random source at the sample size 
``1000 of 2GB-long sequences'', it is expected to 
have a statistical distance smaller than 0.03
and an RMSD smaller than 0.001
for the standard normal distribution $\mu^U_n$ 
(see, e.g., \cite{freedmanstatistics}).
At sample size 10000 of 2GB sequences,
the statistical distance is reduced to 0.02 which is still 
more than acceptable for cryptographic applications.

One could also visually analyze the pictures
in Figure \ref{nistsha1all}.
For example, from three pictures in  Figure \ref{nistsha1all},
we may get the following impression: 
sequences generated by Java SHA1PRNG have a good performance to
stay within the interval $[-1,1]$ though there is a big 
gap between 400 and 900 in the bottom area that is close to the line 
$y=-1$.
Among the three pictures, sequences generated by SHA1-DRBG  have 
a better performance that looks more close to a true random source.
For sequences generated by  SHA2-DRBG, too many sequences reach 
or go beyond $y=1$ and $y=-1$ lines.

\section{Conclusion}
\label{conclusionSec}
This paper proposed statistical distance based LIL testing 
techniques and showed that, at sample size 1000,
the collection of sequences generated by 
several commonly used pseudorandom generators has a statistical distance 
0.06 and root-mean-square deviation 0.005
from a true random source. These values are larger
than expected for various cryptographic applications.
This paper also calculated the probability for weak 
LIL tests on sequences of less than 2GB. For longer sequences, the
corresponding probabilities decrease significantly. Thus large
sample sizes of sequences are needed for better LIL testing.
Alternatively,  one may also split longer sequences
into independent sub-sequences of 2GB each 
and then use the probabilities
calculated in this paper to carry out LIL testing on them. 
For strong LIL tests, this paper obtained a preliminary
result with a very small probability for a random sequence to pass. 
It would be interesting 
to calculate the exact probability ${\bf SP}_{(\alpha, \aleph)}$
for continuous interval $\aleph$ (e.g., $\aleph=[2^{26}, 2^{34}]$).
We believe that  ${\bf SP}_{(\alpha, \aleph)}$ is large enough for a
reasonable interval $\aleph$ such as $\aleph=[2^{26}, 2^{34}]$.
When  the probability ${\bf SP}_{(\alpha, \aleph)}$ becomes larger,
the required sample size for the strong LIL testing will be smaller.
It would  also be very important to find new techniques that could be used
to design pseudorandom generators with smaller
statistical distance and smaller root-mean-square deviation 
from a true random source.

\bibliographystyle{plain}
\ifbibfile
\bibliography{yonggewang}

\begin{thebibliography}{10}

\bibitem{guardianSnowden}
J.~Ball, J.~Borger, and G.~Greenwald.
\newblock Revealed: how {US} and {UK} spy agencies defeat internet privacy and
  security.
\newblock
  http://www.theguardian.com/world/2013/sep/05/nsa-gchq-encryption-codes-security,
  Sept. 13, 2013.

\bibitem{nist80090a}
E.~Barker and J.~Kelsey.
\newblock {\em NIST SP 800-90A: Recommendation for Random Number Generation
  Using Deterministic Random Bit Generators}.
\newblock NIST, 2012.

\bibitem{blum}
M.~Blum and S.~Micali.
\newblock How to generate cryptographically strong sequences of pseudorandom
  bits.
\newblock {\em SIAM J. Comput.}, 13:850--864, 1984.

\bibitem{clarkson1933definitions}
J.A. Clarkson and C.R. Adams.
\newblock On definitions of bounded variation for functions of two variables.
\newblock {\em Tran. AMS}, 35(4):824--854, 1933.

\bibitem{erdos1946certain}
P.~Erd{\"o}s and M.~Kac.
\newblock On certain limit theorems of the theory of probability.
\newblock {\em Bulletin of AMS}, 52(4):292--302, 1946.

\bibitem{feller1945fundamental}
W.~Feller.
\newblock The fundamental limit theorems in probability.
\newblock {\em Bulletin of AMS}, 51(11):800--832, 1945.

\bibitem{feller}
W.~Feller.
\newblock {\em Introduction to probability theory and its applicatons},
  volume~I.
\newblock John Wiley $\&$ Sons, Inc., New York, 1968.

\bibitem{freedmanstatistics}
D.~Freedman, R.~Pisani, and R.~Purves.
\newblock {\em Statistics}.
\newblock Norton \& Company, 2007.

\bibitem{goldreich2005foundations}
O.~Goldreich.
\newblock {\em Foundations of cryptography: a primer}.
\newblock Now Publishers Inc, 2005.

\bibitem{goldwasser1984probabilistic}
S.~Goldwasser and S.~Micali.
\newblock Probabilistic encryption.
\newblock {\em J. Comput. Sys. Sci.}, 28(2):270--299, 1984.

\bibitem{hellinger1909neue}
E.~Hellinger.
\newblock Neue begr{\"u}ndung der theorie quadratischer formen von
  unendlichvielen ver{\"a}nderlichen.
\newblock {\em J. f{\"u}r die reine und angewandte Mathematik}, 136:210--271,
  1909.

\bibitem{khintchine1924satz}
A.~Khintchine.
\newblock {\"U}ber einen satz der wahrscheinlichkeitsrechnung.
\newblock {\em Fund. Math}, 6:9--20, 1924.

\bibitem{kolmogorov}
A.~N. Kolmogorov.
\newblock Three approaches to the definition of the concept ``quantity of
  information".
\newblock {\em Problems Inform. Transmission}, 1:3--7, 1965.

\bibitem{martin}
P.~Martin-L\"{o}f.
\newblock The definition of random sequences.
\newblock {\em Inform. and Control}, 9:602--619, 1966.

\bibitem{nistrngtool}
NIST.
\newblock Test suite, \url{http://csrc.nist.gov/groups/ST/toolkit/rng/}, 2010.

\bibitem{nytSnowden}
N.~Perlroth, J.~Larson, and S.~Shane.
\newblock {NSA} able to foil basic safeguards of privacy on web.
\newblock
  http://www.nytimes.com/2013/09/06/us/nsa-foils-much-internet-encryption.html,
  Sep. 5, 2013.

\bibitem{nist80022}
A.~Rukhin, J.~Soto, J.~Nechvatal, M.~Smid, E.~Barker, S.~Leigh, M.~Levenson,
  M.~Vangel, D.~Banks, A.~Heckert, J.~Dray, and S.~Vo.
\newblock {\em A Statistical Test Suite for Random and Pseudorandom Number
  Generators for Cryptographic Applications}.
\newblock NIST SP 800-22, 2010.

\bibitem{schnorr1}
C.~P. Schnorr.
\newblock {\em Zuf\"{a}lligkeit und Wahrscheinlichkeit}.
\newblock Lecture Notes in Math. 218. Springer Verlag, 1971.

\bibitem{ville}
J.~Ville.
\newblock {\em \'{E}tude Critique de la Notion de Collectif}.
\newblock Gauthiers-Villars, Paris, 1939.

\bibitem{mises}
R.~von Mises.
\newblock Grundlagen der wahrscheinlichkeitsrechung.
\newblock {\em Math. Z.}, 5:52--89, 1919.

\bibitem{DBLPWang96}
Yongge Wang.
\newblock The law of the iterated logarithm for p-random sequences.
\newblock In {\em IEEE Conf. Comput. Complexity}, pages 180--189, 1996.

\bibitem{wangThesis}
Yongge Wang.
\newblock Randomness and complexity.
\newblock {\em PhD Thesis, University of Heidelberg}, 1996.

\bibitem{wanginfo5}
Yongge Wang.
\newblock Resource bounded randomness and computational complexity.
\newblock {\em Theoret. Comput. Sci.}, 237:33--55, 2000.

\bibitem{wang2002comparison}
Yongge Wang.
\newblock A comparison of two approaches to pseudorandomness.
\newblock {\em Theoretical computer science}, 276(1):449--459, 2002.

\bibitem{yao}
A.~C. Yao.
\newblock Theory and applications of trapdoor functions.
\newblock In {\em Proc. 23rd IEEE FOCS}, pages 80--91, 1982.

\end{thebibliography}
\fi

\section{Appendix}
\label{appendixsec}

\begin{figure}[htb]
\caption{Density functions for distributions  $\mu_n^U$ with $n=2^{26}, \cdots, 2^{34}$}
\label{snapShotLIL}
\begin{center}
\includegraphics[width=0.45\textwidth]{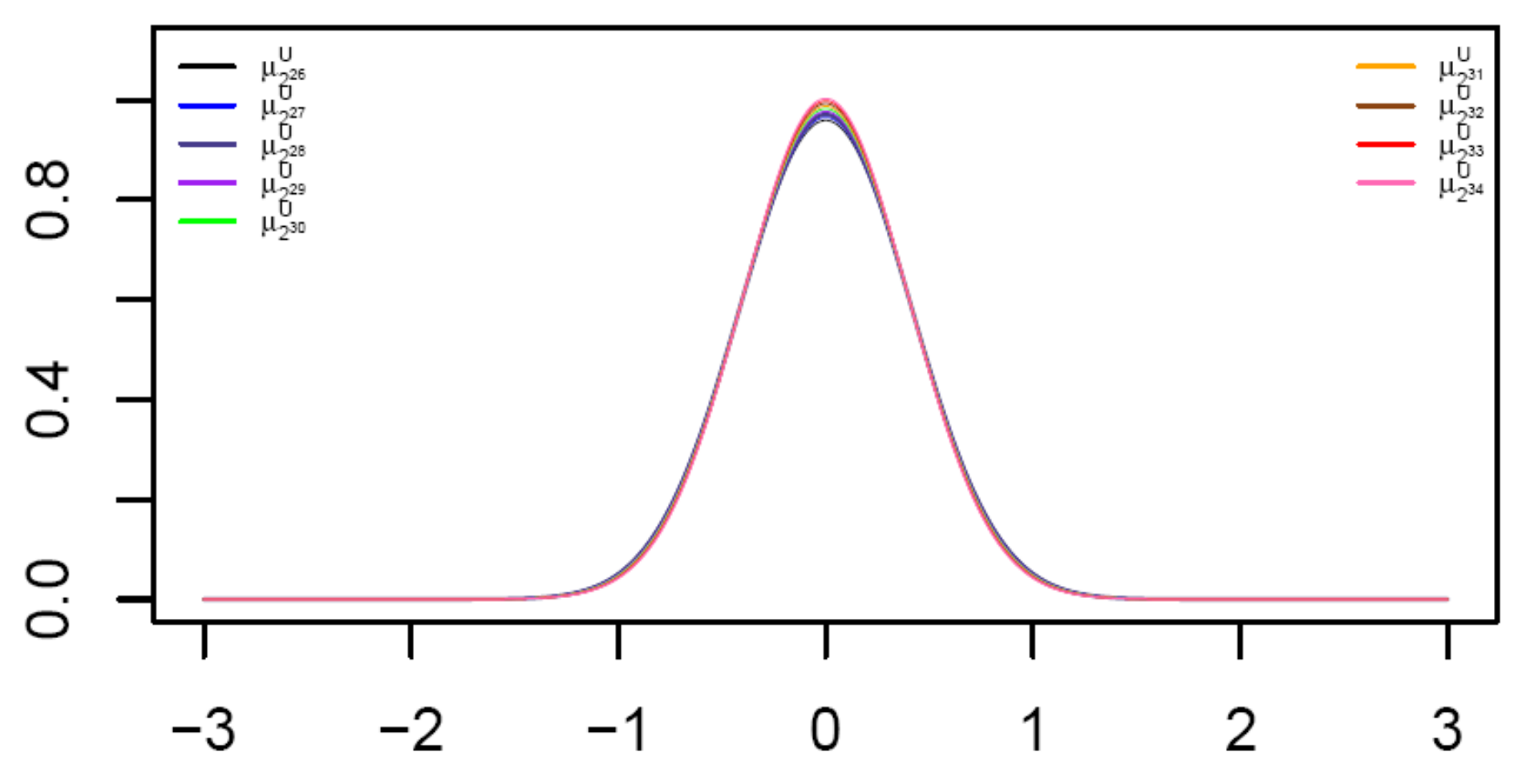} 
\end{center}
\end{figure}

Figure \ref{snapShotLIL}
shows the distributions of  $\mu_n^U$ for $n=2^{26}, \cdots, 2^{34}$
and Table \ref{inducedProbslil} lists values  $\mu_n^U(I)$  on ${\cal B}$  
with $n=2^{26}, \cdots, 2^{34}$. Since $\mu_n^U(I)$ is symmetric,
it is sufficient to list the distribution in the positive side of
the real line.
\begin{table*}[htb]
\caption{The distribution $\mu_n^U$ induced by $S_{lil}$ for 
$n=2^{26},\cdots, 2^{34}$ (due to symmetry, only distribution 
on the positive part of real line $R$ is given)}
\label{inducedProbslil}
\begin{center}
{
\scriptsize
\begin{tabular}{|r|c|c|c|c|c|c|c|c|c|} \hline
 &$2^{26}$ & $2^{27}$ & $2^{28}$ & $2^{29}$ & $2^{30}$ & 
     $2^{31}$&  $2^{32}$ &  $2^{33}$ &  $2^{34}$ \\ \hline
$[0.00,0.05)$   &.047854&.048164&.048460&.048745&.049018&.049281&.049534&.049778&.050013 \\ \hline
$[0.05,0.10)$   &.047168&.047464&.047748&.048020&.048281&.048532&.048773&.049006&.049230\\ \hline
$[0.10,0.15)$   &.045825&.046096&.046354&0.04660&.046839&.047067&.047287&.047498&.047701\\ \hline
$[0.15,0.20)$   &.043882&.044116&.044340&.044553&.044758&.044953&.045141&.045322&.045496\\ \hline
$[0.20,0.25)$   &.041419&.041609&.041789&.041961&.042125&.042282&.042432&.042575&.042713\\ \hline
$[0.25,0.30)$   &.038534&.038674&.038807&.038932&.039051&.039164&.039272&.039375&.039473\\ \hline
$[0.30,0.35)$   &.035336&.035424&.035507&.035584&.035657&.035725&0.03579&.035850&.035907\\ \hline
$[0.35,0.40)$   &.031939&.031976&.032010&.032041&.032068&.032093&.032115&.032135&.032153\\ \hline
$[0.40,0.45)$   &.028454&.028445&.028434&.028421&.028407&.028392&.028375&.028358&.028340\\ \hline
$[0.45,0.50)$   &.024986&.024936&.024886&.024835&.024785&.024735&.024686&.024637&.024588\\ \hline
$[0.50,0.55)$   &.021627&.021542&.021460&.021379&.021300&.021222&.021146&.021072&.020999\\ \hline
$[0.55,0.60)$   &.018450&.018340&.018234&.018130&.018029&.017931&.017836&.017743&.017653\\ \hline
$[0.60,0.65)$   &.015515&.015388&.015265&.015146&.015032&.014921&.014813&.014709&.014608\\ \hline
$[0.65,0.70)$   &.012859&.012723&.012591&.012465&.012344&.012227&.012114&.012004&.011899\\ \hline
$[0.70,0.75)$   &.010506&.010367&.010234&.010106&.009984&.009867&.009754&.009645&.009541\\ \hline
$[0.75,0.80)$   &.008460&.008324&.008195&.008072&.007954&.007841&.007733&.007629&.007530\\ \hline
$[0.80,0.85)$   &.006714&.006587&.006466&.006351&.006241&.006137&.006037&.005941&.005850 \\ \hline
$[0.85,0.90)$   &.005253&.005137&.005027&.004923&.004824&.004730&.004640&.004555&.004474 \\ \hline
$[0.90,0.95)$   &.004050&.003948&.003851&.003759&.003672&.003590&.003512&.003438&.003368 \\ \hline
$[0.95,1.00)$   &.003079&.002990&.002906&.002828&.002754&.002684&.002617&.002555&.002495 \\ \hline
$[1.00,\infty)$ &.008090&.007750&.007437&.007147&.006877&.006627&.006393&.006175&.005970 \\ \hline
\end{tabular}
}
\end{center}
\end{table*}

Table \ref{inducedProbieee} lists values  $\mu_n^{JavaSHA1}(I)$ 
on ${\cal B}$  with $n=2^{26}, \cdots, 2^{34}$.
\begin{table*}[htb]
\caption{The distribution $\mu_n^{JavaSHA1}$ induced by $S_{lil}$  for 
$n=2^{26},\cdots, 2^{34}$}
\label{inducedProbieee}
\begin{center}
{
\scriptsize
\begin{tabular}{|r|c|c|c|c|c|c|c|c|c|} \hline
 &$2^{26}$ & $2^{27}$ & $2^{28}$ & $2^{29}$ & $2^{30}$ & 
     $2^{31}$&  $2^{32}$ &  $2^{33}$ &  $2^{34}$ \\ \hline
$(-\infty,-1)$ &.011&.008&.012&.007&.006&.006&.008&.006&.004\\ \hline
$[-0.1,-0.95)$ &.002&.005&.002&.002&.004&.001&.001&.002&.002\\ \hline
$[-0.95,-0.90)$&.005&.007&.004&.008&.004&.004&.003&.003&.003\\ \hline
$[-0.90,-0.85)$&.008&.005&.006&.003&.008&.005&.001&.003&.007\\ \hline
$[-0.85,-0.80)$&.007&.011&.006&.005&.007&.006&.003&.004&.006\\ \hline
$[-0.80,-0.75)$&.010&.006&.010&.011&.010&.005&.003&.008&.006\\ \hline
$[-0.75,-0.70)$&.015&.010&.013&.010&.002&.004&.013&.011&.012\\ \hline
$[-0.70,-0.65)$&.013&.017&.010&.007&.010&.006&.011&.009&.009\\ \hline
$[-0.65,-0.60)$&.019&.017&.013&.013&.011&.017&.011&.013&.007\\ \hline
$[-0.60,-0.55)$&.014&.021&.015&.022&.019&.018&.017&.022&.017\\ \hline
$[-0.55,-0.50)$&.020&.032&.024&.019&.022&.022&.021&.021&.020\\ \hline
$[-0.50,-0.45)$&.030&.030&.027&.028&.024&.022&.027&.025&.022\\ \hline
$[-0.45,-0.40)$&.034&.035&.037&.021&.025&.020&.031&.033&.037\\ \hline
$[-0.40,-0.35)$&.036&.035&.037&.038&.033&.037&.032&.039&.032\\ \hline
$[-0.35,-0.30)$&.042&.037&.044&.031&.034&.035&.035&.033&.042\\ \hline
$[-0.30,-0.25)$&.043&.033&.042&.039&.032&.043&.046&.040&.041\\ \hline
$[-0.25,-0.20)$&.042&.039&.040&.053&.048&.039&.047&.039&.048\\ \hline
$[-0.20,-0.15)$&.053&.047&.042&.049&.052&.042&.039&.038&.029\\ \hline
$[-0.15,-0.10)$&.055&.045&.049&.056&.053&.038&.048&.052&.043\\ \hline
$[-0.10,-.05)$&.047&.046&.051&.049&.046&.054&.041&.049&.053\\ \hline
$[-.05,0)$    &.040&.037&.048&.047&.045&.055&.053&.059&.048\\ \hline
$[0,.05)$     &.042&.046&.050&.053&.041&.041&.041&.045&.044\\ \hline
$[.05,0.10)$  &.039&.053&.048&.048&.043&.050&.049&.038&.049\\ \hline
$[0.10,0.15)$  &.040&.054&.039&.049&.058&.064&.039&.050&.054\\ \hline
$[0.15,0.20)$  &.042&.047&.039&.047&.051&.058&.064&.041&.038\\ \hline
$[0.20,0.25)$  &.034&.030&.029&.031&.040&.053&.050&.049&.040\\ \hline
$[0.25,0.30)$  &.027&.036&.040&.032&.041&.033&.039&.040&.044\\ \hline
$[0.30,0.35)$  &.034&.027&.034&.033&.043&.022&.033&.040&.040\\ \hline
$[0.35,0.40)$  &.026&.033&.030&.043&.030&.030&.030&.022&.038\\ \hline
$[0.40,0.45)$  &.030&.030&.016&.024&.030&.026&.034&.022&.031\\ \hline
$[0.45,0.50)$  &.020&.021&.023&.028&.019&.033&.028&.022&.021\\ \hline
$[0.50,0.55)$  &.020&.018&.018&.008&.025&.024&.013&.026&.018\\ \hline
$[0.55,0.60)$  &.019&.012&.020&.020&.017&.020&.022&.015&.023\\ \hline
$[0.60,0.65)$  &.015&.015&.014&.009&.015&.015&.015&.017&.019\\ \hline
$[0.65,0.70)$  &.011&.013&.014&.008&.010&.008&.009&.015&.013\\ \hline
$[0.70,0.75)$  &.009&.005&.011&.013&.008&.009&.009&.015&.012\\ \hline
$[0.75,0.80)$  &.011&.009&.007&.004&.006&.009&.009&.006&.003\\ \hline
$[0.80,0.85)$  &.007&.008&.009&.004&.008&.009&.002&.009&.007\\ \hline
$[0.85,0.90)$  &.008&.004&.007&.008&.004&.003&.006&.008&.007\\ \hline
$[0.90,0.95)$  &.006&.004&.007&.002&.004&.004&.002&.004&.003\\ \hline
$[0.95,1.00)$  &.003&.004&.002&.010&.002&.004&.004&.002&.002\\ \hline
$[1.00,\infty)$&.011&.008&.011&.008&.010&.006&.011&.005&.006\\ \hline
\end{tabular}
}
\end{center}
\end{table*}

Table \ref{inducedProbnistSHA1} lists values  
$\mu_n^{nistDRBsha1}(I)$ on ${\cal B}$ with $n=2^{26}, \cdots, 2^{34}$.
\begin{table*}[htb]
\caption{The distribution $\mu_n^{nistDRBGsha1}$ induced by $S_{lil}$  for 
$n=2^{26},\cdots, 2^{34}$}
\label{inducedProbnistSHA1}
\begin{center}
{
\scriptsize
\begin{tabular}{|r|c|c|c|c|c|c|c|c|c|} \hline
 &$2^{26}$ & $2^{27}$ & $2^{28}$ & $2^{29}$ & $2^{30}$ & 
     $2^{31}$&  $2^{32}$ &  $2^{33}$ &  $2^{34}$ \\ \hline
$(-\infty,-1)$ &.009&.008&.007&.008&.006&.007&.007&.006&.007\\ \hline
$[-0.1,-0.95)$ &.002&.004&.001&.005&.002&.003&.003&.001&.005 \\ \hline
$[-0.95,-0.90)$&.004&.007&.004&.005&.002&.006&.004&.002&.000\\ \hline
$[-0.90,-0.85)$&.009&.006&.011&.008&.005&.003&.006&.006&.009\\ \hline
$[-0.85,-0.80)$&.005&.010&.004&.010&.008&.003&.004&.010&.003\\ \hline
$[-0.80,-0.75)$&.007&.004&.010&.011&.006&.008&.011&.005&.002\\ \hline
$[-0.75,-0.70)$&.009&.005&.014&.008&.011&.017&.007&.013&.011\\ \hline
$[-0.70,-0.65)$&.019&.014&.014&.011&.026&.015&.012&.013&.009\\ \hline
$[-0.65,-0.60)$&.013&.020&.010&.012&.018&.011&.014&.012&.011\\ \hline
$[-0.60,-0.55)$&.016&.021&.019&.014&.019&.022&.021&.018&.017\\ \hline
$[-0.55,-0.50)$&.022&.018&.022&.027&.028&.022&.023&.023&.023\\ \hline
$[-0.50,-0.45)$&.027&.025&.020&.033&.021&.029&.025&.026&.034\\ \hline
$[-0.45,-0.40)$&.028&.030&.024&.027&.025&.033&.034&.028&.035\\ \hline
$[-0.40,-0.35)$&.030&.036&.031&.026&.027&.026&.037&.041&.036\\ \hline
$[-0.35,-0.30)$&.041&.032&.037&.035&.032&.026&.040&.039&.038\\ \hline
$[-0.30,-0.25)$&.034&.043&.052&.038&.039&.032&.034&.032&.048\\ \hline
$[-0.25,-0.20)$&.045&.031&.048&.038&.038&.046&.036&.030&.044\\ \hline
$[-0.20,-0.15)$&.055&.044&.048&.039&.039&.042&.046&.051&.050\\ \hline
$[-0.15,-0.10)$&.056&.058&.046&.046&.041&.050&.046&.050&.042\\ \hline
$[-0.10,-0.05)$&.046&.048&.048&.044&.044&.051&.046&.059&.039\\ \hline
$[-0.05,0)$    &.045&.050&.035&.051&.040&.053&.048&.059&.048\\ \hline
$[0,0.05)$     &.045&.040&.051&.052&.047&.041&.033&.044&.042\\ \hline
$[0.05,0.10)$  &.058&.038&.060&.047&.056&.044&.044&.056&.051\\ \hline
$[0.10,0.15)$  &.042&.044&.035&.041&.057&.047&.050&.040&.048\\ \hline
$[0.15,0.20)$  &.037&.040&.040&.051&.039&.049&.045&.038&.033\\ \hline
$[0.20,0.25)$  &.034&.050&.037&.056&.045&.039&.046&.039&.033\\ \hline
$[0.25,0.30)$  &.042&.041&.034&.046&.042&.032&.037&.039&.035\\ \hline
$[0.30,0.35)$  &.036&.036&.040&.035&.036&.031&.043&.037&.040\\ \hline
$[0.35,0.40)$  &.022&.038&.028&.033&.045&.029&.043&.032&.038\\ \hline
$[0.40,0.45)$  &.029&.020&.026&.023&.037&.036&.031&.018&.034\\ \hline
$[0.45,0.50)$  &.025&.026&.028&.023&.019&.029&.020&.019&.026\\ \hline
$[0.50,0.55)$  &.024&.025&.034&.019&.012&.031&.024&.023&.031\\ \hline
$[0.55,0.60)$  &.020&.012&.016&.015&.023&.020&.019&.022&.014\\ \hline
$[0.60,0.65)$  &.010&.016&.011&.014&.013&.019&.011&.011&.015\\ \hline
$[0.65,0.70)$  &.012&.013&.011&.008&.015&.012&.010&.013&.013\\ \hline
$[0.70,0.75)$  &.006&.012&.011&.008&.012&.011&.011&.014&.006\\ \hline
$[0.75,0.80)$  &.010&.011&.005&.012&.009&.006&.009&.006&.011\\ \hline
$[0.80,0.85)$  &.006&.005&.006&.005&.006&.005&.002&.008&.006\\ \hline
$[0.85,0.90)$  &.005&.003&.006&.003&.002&.005&.001&.007&.005\\ \hline
$[0.90,0.95)$  &.005&.007&.003&.002&.003&.004&.006&.004&.002\\ \hline
$[0.95,1.00)$  &.002&.004&.003&.004&.001&.001&.003&.001&.001\\ \hline
$[1.00,\infty)$&.008&.005&.010&.007&.004&.004&.008&.005&.005\\ \hline
\end{tabular}
}
\end{center}
\end{table*}

Table \ref{inducedProbnistSHA256} lists values  
$\mu_n^{nistDRBGsha256,1000}(I)$ on ${\cal B}$ with $n=2^{26}, \cdots, 2^{34}$.
\begin{table*}[htb]
\caption{The distribution $\mu_n^{nistDRBGsha256,1000}$ induced by $S_{lil}$  for 
$n=2^{26},\cdots, 2^{34}$}
\label{inducedProbnistSHA256}
\begin{center}
{
\scriptsize
\begin{tabular}{|r|c|c|c|c|c|c|c|c|c|} \hline
 &$2^{26}$ & $2^{27}$ & $2^{28}$ & $2^{29}$ & $2^{30}$ & 
     $2^{31}$&  $2^{32}$ &  $2^{33}$ &  $2^{34}$ \\ \hline
$(-\infty,-1)$ &.007&.005&.005&.002&.004&.003&.003&.009&.006 \\ \hline
$[-0.1,-0.95)$ &.006&.004&.003&.002&.004&.003&.006&.003&.003 \\ \hline
$[-0.95,-0.90)$&.003&.004&.006&.001&.005&.003&.002&.001&.001 \\ \hline
$[-0.90,-0.85)$&.004&.006&.003&.005&.004&.005&.002&.005&.003\\ \hline
$[-0.85,-0.80)$&.007&.006&.002&.013&.005&.007&.011&.005&.004\\ \hline
$[-0.80,-0.75)$&.008&.010&.007&.006&.004&.008&.013&.007&.004\\ \hline
$[-0.75,-0.70)$&.007&.010&.010&.013&.005&.004&.009&.010&.006\\ \hline
$[-0.70,-0.65)$&.021&.013&.012&.015&.006&.018&.011&.010&.008\\ \hline
$[-0.65,-0.60)$&.009&.008&.012&.015&.021&.009&.014&.019&.022\\ \hline
$[-0.60,-0.55)$&.016&.019&.019&.018&.016&.008&.020&.012&.015\\ \hline
$[-0.55,-0.50)$&.025&.013&.021&.016&.017&.023&.021&.013&.020\\ \hline
$[-0.50,-0.45)$&.014&.033&.026&.023&.018&.015&.025&.034&.025\\ \hline
$[-0.45,-0.40)$&.028&.024&.033&.023&.034&.034&.030&.026&.022\\ \hline
$[-0.40,-0.35)$&.021&.025&.031&.034&.029&.036&.032&.033&.022\\ \hline
$[-0.35,-0.30)$&.034&.031&.039&.043&.037&.040&.024&.031&.037\\ \hline
$[-0.30,-0.25)$&.042&.041&.036&.027&.033&.031&.036&.041&.036\\ \hline
$[-0.25,-0.20)$&.043&.046&.035&.030&.045&.039&.039&.037&.042\\ \hline
$[-0.20,-0.15)$&.040&.042&.051&.047&.042&.044&.036&.042&.046\\ \hline
$[-0.15,-0.10)$&.039&.042&.038&.050&.055&.044&.053&.043&.046\\ \hline
$[-0.10,-0.05)$&.048&.046&.042&.055&.045&.050&.045&.042&.049\\ \hline
$[-0.05,0)$    &.049&.045&.044&.043&.045&.049&.040&.063&.055\\ \hline
$[0,0.05)$     &.055&.059&.050&.062&.049&.054&.056&.040&.043\\ \hline
$[0.05,0.10)$  &.043&.041&.049&.044&.049&.045&.059&.060&.047\\ \hline
$[0.10,0.15)$  &.046&.045&.036&.038&.045&.045&.042&.052&.052\\ \hline
$[0.15,0.20)$  &.049&.046&.052&.040&.045&.049&.048&.047&.050\\ \hline
$[0.20,0.25)$  &.054&.043&.033&.046&.046&.047&.033&.037&.043\\ \hline
$[0.25,0.30)$  &.044&.050&.046&.041&.052&.039&.038&.040&.047\\ \hline
$[0.30,0.35)$  &.037&.030&.032&.033&.035&.037&.034&.036&.054\\ \hline
$[0.35,0.40)$  &.033&.028&.030&.040&.039&.033&.036&.049&.032\\ \hline
$[0.40,0.45)$  &.025&.030&.036&.027&.024&.026&.029&.025&.033\\ \hline
$[0.45,0.50)$  &.022&.031&.025&.043&.025&.032&.027&.028&.022\\ \hline
$[0.50,0.55)$  &.023&.026&.021&.016&.027&.023&.018&.019&.020\\ \hline
$[0.55,0.60)$  &.017&.017&.020&.012&.019&.017&.028&.020&.019\\ \hline
$[0.60,0.65)$  &.024&.016&.018&.014&.025&.022&.018&.011&.015\\ \hline
$[0.65,0.70)$  &.008&.016&.017&.009&.013&.017&.014&.007&.012\\ \hline
$[0.70,0.75)$  &.013&.007&.016&.014&.006&.007&.014&.008&.016\\ \hline
$[0.75,0.80)$  &.002&.009&.011&.010&.009&.011&.004&.008&.004\\ \hline
$[0.80,0.85)$  &.011&.011&.012&.007&.001&.004&.005&.007&.007\\ \hline
$[0.85,0.90)$  &.010&.006&.007&.003&.004&.004&.004&.004&.003\\ \hline
$[0.90,0.95)$  &.004&.006&.002&.005&.004&.005&.005&.002&.006\\ \hline
$[0.95,1.00)$  &.002&.003&.002&.007&.001&.002&.005&.003&.000\\ \hline
$[1.00,\infty)$&.007&.007&.010&.008&.008&.008&.011&.011&.003\\ \hline
\end{tabular}
}
\end{center}
\end{table*}

\begin{figure}[htb]
\caption{The distributions $\mu_n^U$ and  $\mu_n^{nistDRBGsha256,1000}$
with $n=2^{26}, \cdots, 2^{34}$}
\label{snapShotnistSHA256}
\begin{center}
\includegraphics[width=0.45\textwidth]{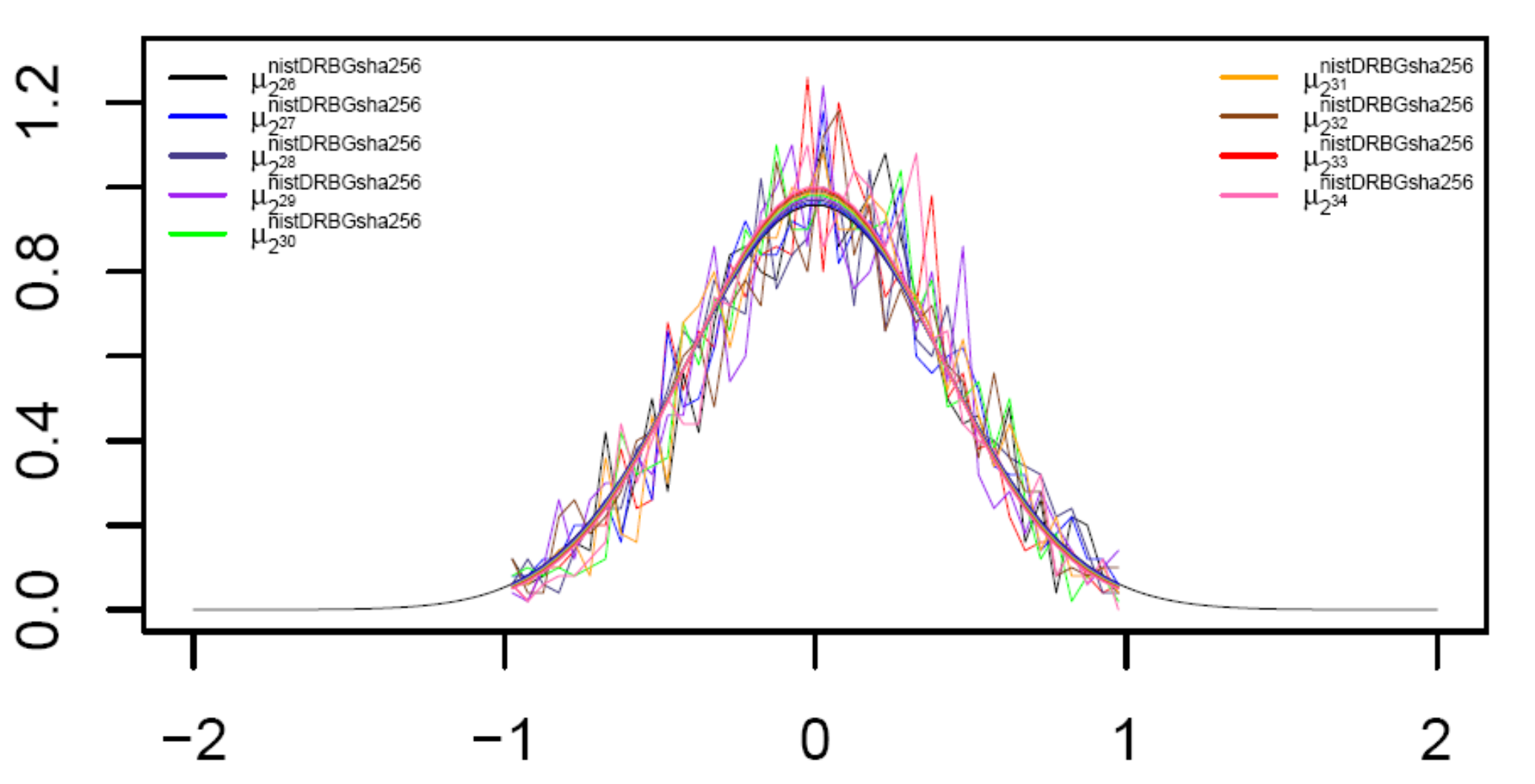} 
\end{center}
\end{figure}

Figure \ref{snapShotnistSHA256} compared the distributions
$\mu_n^{nistDRBGsha256,1000}$.

Table \ref{inducedProbnistSHA256} lists values  
$\mu_n^{nistDRBGsha256,10000}(I)$ on ${\cal B}$ with $n=2^{26}, \cdots, 2^{34}$.
\begin{table*}[htb]
\caption{The distribution $\mu_n^{nistDRBGsha256,10000}$ induced by $S_{lil}$  for 
$n=2^{26},\cdots, 2^{34}$}
\label{inducedProbnistSHA25610k}
\begin{center}
{
\scriptsize
\begin{tabular}{|r|c|c|c|c|c|c|c|c|c|} \hline
 &$2^{26}$ & $2^{27}$ & $2^{28}$ & $2^{29}$ & $2^{30}$ & 
     $2^{31}$&  $2^{32}$ &  $2^{33}$ &  $2^{34}$ \\ \hline
$(-\infty,-1)$ &.0071&.0070&.0062&.0067&.0061&.0066&.0069&.0053&.0055\\ \hline
$[-0.1,-0.95)$ &.0036&.0036&.0030&.0032&.0030&.0027&.0026&.0031&.0023\\ \hline
$[-0.95,-0.90)$&.0047&.0036&.0050&.0031&.0032&.0035&.0028&.0036&.0029\\ \hline
$[-0.90,-0.85)$&.0044&.0057&.0060&.0035&.0039&.0047&.0038&.0043&.0035\\ \hline
$[-0.85,-0.80)$&.0063&.0068&.0058&.0085&.0057&.0062&.0066&.0062&.0050\\ \hline
$[-0.80,-0.75)$&.0089&.0078&.0090&.0082&.0071&.0057&.0083&.0071&.0070\\ \hline
$[-0.75,-0.70)$&.0112&.0102&.0103&.0094&.0096&.0097&.0108&.0081&.0099\\ \hline
$[-0.70,-0.65)$&.0126&.0128&.0118&.0118&.0118&.0113&.0104&.0123&.0120\\ \hline
$[-0.65,-0.60)$&.0149&.0147&.0166&.0166&.0151&.0147&.0185&.0144&.0147\\ \hline
$[-0.60,-0.55)$&.0180&.0217&.0179&.0181&.0191&.0180&.0165&.0169&.0199\\ \hline
$[-0.55,-0.50)$&.0216&.0197&.0215&.0217&.0201&.0247&.0243&.0186&.0188\\ \hline
$[-0.50,-0.45)$&.0228&.0275&.0245&.0228&.0226&.0220&.0250&.0246&.0255\\ \hline
$[-0.45,-0.40)$&.0274&.0303&.0310&.0309&.0292&.0283&.0319&.0302&.0287\\ \hline
$[-0.40,-0.35)$&.0302&.0298&.0322&.0331&.0315&.0326&.0323&.0354&.0336\\ \hline
$[-0.35,-0.30)$&.0353&.0346&.0344&.0341&.0361&.0385&.0331&.0361&.0329\\ \hline
$[-0.30,-0.25)$&.0394&.0385&.0365&.0379&.0391&.0408&.0381&.0375&.0387\\ \hline
$[-0.25,-0.20)$&.0435&.0405&.0391&.0425&.0462&.0375&.0454&.0442&.0446\\ \hline
$[-0.20,-0.15)$&.0419&.0436&.0430&.0430&.0450&.0488&.0431&.0429&.0453\\ \hline
$[-0.15,-0.10)$&.0439&.0475&.0446&.0475&.0506&.0450&.0464&.0466&.0491\\ \hline
$[-0.10,-0.05)$&.0474&.0426&.0516&.0484&.0480&.0499&.0474&.0511&.0501\\ \hline
$[-0.05,0)$    &.0488&.0489&.0473&.0447&.0474&.0471&.0465&.0501&.0481\\ \hline
$[0,0.05)$     &.0497&.0478&.0499&.0460&.0499&.0505&.0495&.0507&.0485\\ \hline
$[0.05,0.10)$  &.0466&.0460&.0470&.0493&.0512&.0465&.0474&.0476&.0469\\ \hline
$[0.10,0.15)$  &.0436&.0478&.0479&.0455&.0475&.0481&.0466&.0468&.0494\\ \hline
$[0.15,0.20)$  &.0450&.0455&.0467&.0438&.0436&.0459&.0487&.0472&.0469\\ \hline
$[0.20,0.25)$  &.0435&.0411&.0389&.0440&.0418&.0466&.0407&.0460&.0431\\ \hline
$[0.25,0.30)$  &.0393&.0395&.0392&.0406&.0414&.0390&.0407&.0381&.0405\\ \hline
$[0.30,0.35)$  &.0370&.0351&.0325&.0377&.0334&.0341&.0357&.0348&.0352\\ \hline
$[0.35,0.40)$  &.0319&.0304&.0323&.0321&.0289&.0300&.0290&.0363&.0347\\ \hline
$[0.40,0.45)$  &.0308&.0286&.0295&.0309&.0264&.0274&.0271&.0300&.0293\\ \hline
$[0.45,0.50)$  &.0239&.0235&.0249&.0252&.0251&.0243&.0243&.0241&.0257\\ \hline
$[0.50,0.55)$  &.0203&.0229&.0184&.0219&.0213&.0226&.0219&.0201&.0202\\ \hline
$[0.55,0.60)$  &.0166&.0177&.0166&.0154&.0192&.0168&.0189&.0158&.0178\\ \hline
$[0.60,0.65)$  &.0162&.0150&.0160&.0163&.0167&.0154&.0138&.0127&.0144\\ \hline
$[0.65,0.70)$  &.0137&.0143&.0145&.0119&.0120&.0122&.0123&.0123&.0111\\ \hline
$[0.70,0.75)$  &.0102&.0103&.0111&.0092&.0109&.0103&.0104&.0088&.0091\\ \hline
$[0.75,0.80)$  &.0074&.0087&.0089&.0074&.0082&.0079&.0084&.0080&.0070\\ \hline
$[0.80,0.85)$  &.0081&.0070&.0075&.0068&.0060&.0063&.0069&.0050&.0067\\ \hline
$[0.85,0.90)$  &.0059&.0057&.0047&.0058&.0033&.0050&.0037&.0050&.0040\\ \hline
$[0.90,0.95)$  &.0044&.0050&.0035&.0035&.0039&.0035&.0040&.0037&.0044\\ \hline
$[0.95,1.00)$  &.0032&.0037&.0033&.0024&.0021&.0027&.0026&.0023&.0015\\ \hline
$[1.00,\infty)$&.0088&.0070&.0094&.0086&.0068&.0066&.0067&.0061&.0055\\ \hline
\end{tabular}
}
\end{center}
\end{table*}

\begin{figure}[htb]
\caption{The distributions $\mu_n^U$ and  $\mu_n^{nistDRBGsha256,10000}$
with $n=2^{26}, \cdots, 2^{34}$}
\label{snapShotnistSHA25610k}
\begin{center}
\includegraphics[width=0.45\textwidth]{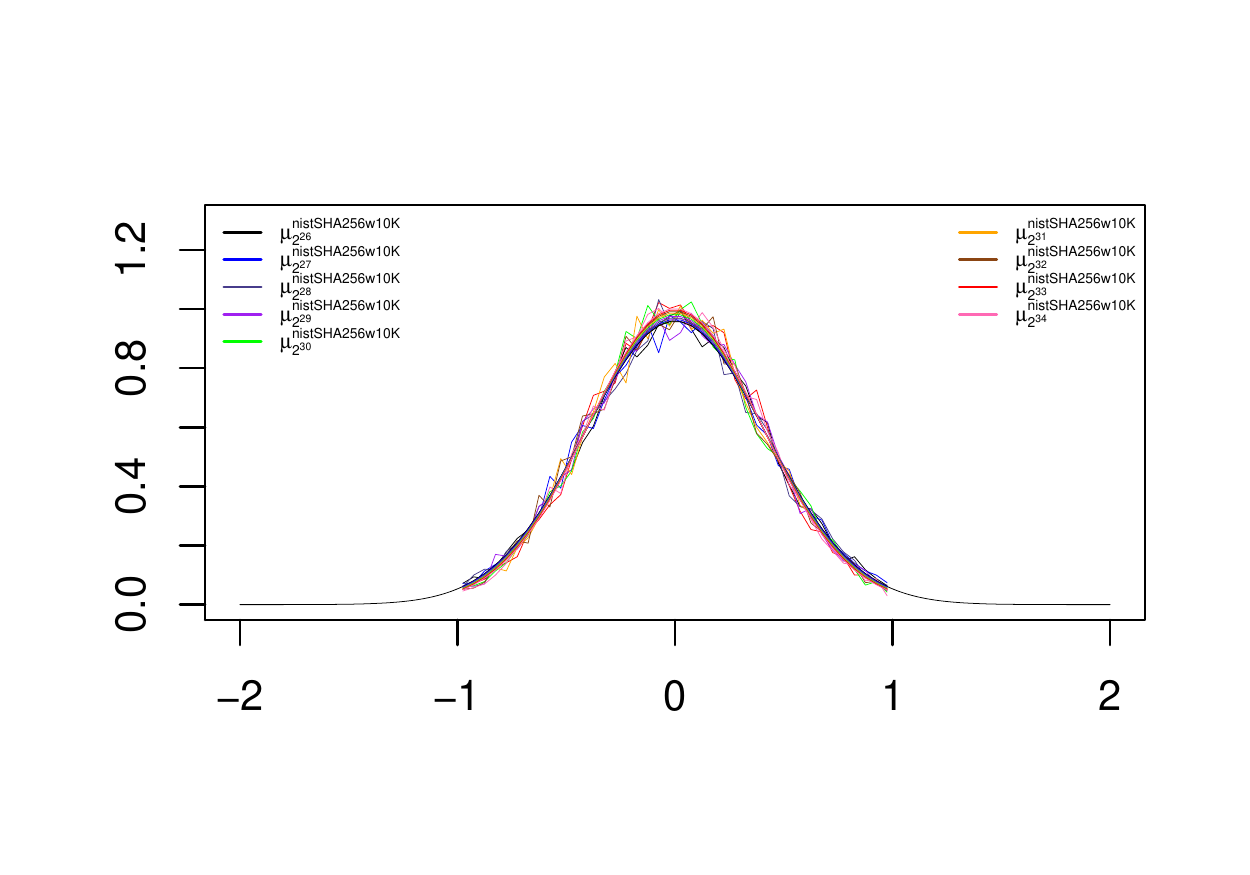} 
\end{center}
\end{figure}

Figure \ref{snapShotnistSHA25610k} compared the distributions
$\mu_n^{nistDRBGsha256}$.

\end{document}